\documentstyle[12pt]{article}
\def\be{\begin{equation}}
\def\ee{\end{equation}}
\def\bea{\begin{eqnarray}}
\def\eea{\end{eqnarray}}
\catcode`@11
\def\seceqaa{\@addtoreset{equation}{section}
\def\theequation{A\arabic{equation}}}
\def\seceqbb{\@addtoreset{equation}{section}
\def\theequation{B\arabic{equation}}}
\catcode`@11
\begin{document}
\begin{titlepage}
\begin{center}
{\Large\bf On `Light' Fermions and Proton Stability in `Big Divisor' $D3$/$D7$ Large Volume Compactifications}
\vskip 0.1in { Aalok Misra\footnote{e-mail: aalokfph@iitr.ernet.in} and
Pramod Shukla\footnote{email: pmathdph@iitr.ernet.in}\\
Department of Physics, Indian Institute of Technology,
Roorkee - 247 667, Uttarakhand, India}
\end{center}
\thispagestyle{empty}

\begin{abstract}
Building up on our earlier work \cite{D3_D7_Misra_Shukla,Sparticles_Misra_Shukla}, we show the possibility of generating ``light" fermion mass scales of MeV-GeV range (possibly related to first two generations of quarks/leptons) as well as ${\rm eV}$ (possibly related to first two generations of neutrinos) in type IIB string theory compactified on Swiss-Cheese orientifolds in the presence of a mobile space-time filling $D3-$brane restricted to (in principle) stacks of fluxed $D7$-branes wrapping the ``big" divisor $\Sigma_B$. This part of the paper is an expanded version of the latter half of section {\bf 3} of a published short invited review  \cite{review_2} written up by one of the authors [AM]. Further, we also show that there are no SUSY GUT-type dimension-five operators corresponding to proton decay, as well as estimate the proton lifetime from a SUSY GUT-type four-fermion dimension-six operator to be $10^{61}$ years. Based on GLSM calculations in \cite{D3_D7_Misra_Shukla} for obtaining the geometric K\"{a}hler potential for the ``big divisor", using further the Donaldson's algorithm, we also briefly discuss in the first of the two appendices, obtaining a metric for the Swiss-Cheese Calabi-Yau used, that becomes Ricci flat in the large volume limit.
\end{abstract}
\end{titlepage}


\section{Introduction}
For the past decade or so, a lot of progress have been made in realistic model building in string theory  and  obtaining numbers which could be matched (directly or indirectly) with some experimental data. In this context, embedding of (MS)SM in string theory with realization of its mass spectrum and cosmological model building with realizing de-Sitter solutions, inflation etc. have been among some of the major issues to work on. A very interesting class of models for realistic model building of cosmology as well as phenomenology has been the L(arge) V(olume) S(cenarios) which was developed in the context of type IIB orientifold compactifications \cite{Balaetal2}.

Further, supporting (MS)SM and study of Susy breaking phenomenon along with realizing its low energy matter content from string theoretic origin has been of quite interest. In the usual Higgs mechanism, fermion masses are generated by electroweak symmetry breaking through giving VEVs to Higgs(es). In the context of realizing fermion masses in ${\cal N}=1$ type IIB orientifold compactifications, one has introduce open string moduli and has to know the explicit K\"{a}hler potential and the superpotential for matter fields, which makes the problem more complicated. Although there has been proposals for realizing fermion masses from a superstring inspired model using Higgs-like mechanism \cite{ferm string}, realizing fermion masses in large volume scenarios inspired models has been missing to best of our knowledge. In this paper, we show the possibility of realizing first two generation fermion mass scales in our Swiss-Cheese orientifold setup.

Also, the compelling evidence of non-zero neutrino masses and their mixing has attracted a lot of attention  as it supports the idea why one should think about physics beyond the Standard Model, which is experimentally well tested. Further, the flavor conversion of solar, atmospheric, reactor, and accelerator neutrinos are convincing enough for nonzero masses of neutrinos and their mixing among themselves similar to that of quarks, provides the first evidence of new physics beyond the standard model. This has motivated enormous amount of activities not only towards particle physics side but also towards cosmology side (like dark energy and dark matter studies) which can be found in plenty of review articles \cite{Mohapatra_group}. Although there has been several aspects for theoretical realization of non-zero neutrino masses with its Dirac-type (e.g. see \cite{dirac_neutrino1,dirac_neutrino2,Hamed_Grossman} in the context of warped compactification) as well as Majorana-type origin, however the models with sea-saw mechanism giving small Majorana neutrino masses has been among the most suited ones (see \cite{seasaw_without high scale1,seasaw_without high scale2,neutrino_flatspace,Mohapatra_group,conlon_neutrino} and reference therein). In the usual sea-saw mechanisms, a high intermediate scale of right handed neutrino (where some new physics starts) lying between TeV and GUT scale, is involved. However there has been proposal of sea-saw mechanisms without requirement of high energy scale \cite{Hamed_Grossman,seasaw_without high scale1,seasaw_without high scale2}. Further, neutrino masses in extra dimensions have been first considered in flat space in \cite{neutrino_flatspace} and followed by this, it has been shown in the context of large extra dimensions that small non-zero values of Dirac as well as Majorana neutrino masses could be realized using some intrinsic higher dimension mechanisms \cite{dirac_neutrino1}. In fact, the mysterious high intermediate scale ($10^{11}-10^{15}$ GeV) required in generating small majorana neutrino masses via sea-saw mechanism has a natural geometric origin in the class of large volume models as suggested in \cite{conlon_neutrino}, and as will be explicitly shown in this paper.

The issue of proton stability which is a generic prediction of Grand unified theories, has been a dramatic outcome of Grand unified theories beyond SM. Although proton decay has not been experimentally observed, usually in Grand unified theories which provide an elegant explanation of various issues of real wold physics, the various decay channels are open due to higher dimensional operators violating baryon (B) numbers. However the life time of the proton (in decay channels) studied in various models has been estimated to be quite large (as $\tau_p\sim M_{X}$ with $M_X$ being some high scale) \cite{Prot_Decay_review}. Further, studies of dimension-five and dimension-six operators relevant to proton decay in SUSY GUT as well as String/M theoretic setups, have been of great importance in obtaining estimates on the lifetime of the proton (See \cite{Prot_Decay_review}).

So far, to our knowledge, a {\it single} framework which is able to reproduce the fermionic mass scales relevant to the quarks/leptons as well as the neutrinos and is able to demonstrate proton stability, has been missing and has remained a long-standing problem. It is the aim of this paper to address this problem.

The paper is organized as follows. We first  briefly discuss, in section {\bf 2}, our  type IIB D3/D7 Swiss-Cheese orientifold large volume setup \cite{D3_D7_Misra_Shukla}. Then in section {\bf 3}, we explore the possibility of realizing fermion (the first two-generation leptons/quarks) mass scales of ${\cal O}(MeV-GeV)$ and (first two-generation neutrino-like) $\leq{\cal O}(eV$) masses, the latter via lepton number violating non-renormalizable dimension-five operators in our setup; this also utilizes MSSM/2HDM one-loop RG flow equations. In section {\bf 4}, with the mobile $D3$-brane restricted to the $D7$-brane stack, we show that there are no SUSY GUT-type dimension-five operators pertaining to proton decay and estimate the proton lifetime from a SUSY GUT-type four-fermion dimension-six operator. There are two appendices: in appendix {\bf A} we construct, in the large volume limit, a basis for $H^{1,1}_-$ and outline the constrcution of a geometric K\"{a}hler potential for the Swiss-Cheese Calabi-Yau (for simplicity, close to the big divisor) that becomes Ricci-flat in the large volume limit (the details of this part of the calculations will appear elsewhere \cite{DM}); in appendix {\bf B} we explicitly show that even in the $D3/D7$ setup, it is possible to introduce a K\"{a}hler hierarchy in the divisor volumes.

\section{The Setup}
Let us first briefly describe our D3/D7 Swiss-Cheese setup \cite{D3_D7_Misra_Shukla} which will be used in this letter. It is type IIB compactified on the orientifold of a ``Swiss-Cheese Calabi-Yau" in the LVS limit. The Swiss-Cheese Calabi Yau we are using, is a
projective variety in ${\ WCP}^4[1,1,1,6,9]$:
\begin{equation}
\label{eq:hyp_def}
x_1^{18} + x_2^{18} + x_3^{18} + x_4^3 + x_5^2 - 18\psi \prod_{i=1}^5x_i - 3\phi x_1^6x_2^6x_3^6 = 0,
\end{equation}
\noindent which has two (big and small) divisors $\Sigma_B(x_5=0)$ and $\Sigma_S(x_4=0)$.
The choice of (\ref{eq:hyp_def}) is well motivated as elucidated by the following arguments.

From Sen's orientifold-limit-of-F-theory point of view  corresponding to type IIB compactified on a Calabi-Yau three fold $Z$-orientifold with $O3/O7$ planes, one requires a Calabi-Yau four-fold $X_4$ elliptically fibered (with projection $\pi$) over a 3-fold $B_3(\equiv CY_3-$orientifold)  where $B_3$ could either be a Fano three-fold or an $n$-twisted ${\bf CP}^1$-fibration over ${\bf CP}^2$ such that pull-back of the divisors in $CY_3$ automatically satisfy Witten's unit-arithmetic genus condition  \cite{DDF,denef_LesHouches}. The toric data of $B_3$ consists of five divisors, three of which are pullbacks of three lines in ${\bf CP}^2$ and the other two are sections of the aforementioned fibration. From the point of view of M-theory compactified on $X_4$, the non-perturbative superpotential receives non-zero contributions from $M5$-brane instantons involving wrapping around uplifts {\bf V} to $X_4$ of ``vertical" divisors ($\pi({\rm\bf V})$ is a proper subset of $B_3$) in $B_3$. These vertical divisors are either components of singular fibers or are pull-backs of smooth divisors in $B_3$. There exists a Weierstrass model $\pi_0:{\cal W}\rightarrow B_3$ and its resolution
$\mu: X_4\rightarrow {\cal W}$. For $n=6$ \cite{DDF}, the $CY_4$ will be the resolution of a Weierstrass model with $D_4$ singularity along the first section and an $E_{6/7/8}$ singularity along the second section. {\it The Calabi-Yau three-fold $Z$ then turns out to be a unique Swiss-Cheese Calabi Yau - an elliptic fibration over ${\bf CP}^2$ in ${\bf WCP}^4[1,1,1,6,9]$ given by (\ref{eq:hyp_def}).} We would be assuming an $E_8$-singularity as this corresponds to
 $h^{1,1}_-(CY_3)=h^{2,1}(CY_4)\neq0$\cite{denef_LesHouches} which is what we will be needing and using. The required Calabi-Yau has $h^{1,1}=2, h^{2,1}=272$. The same has a large discrete symmetry group given by $\Gamma={\bf Z}_6\times{\bf Z}_{18}$ (as mentioned in \cite{D3_D7_Misra_Shukla}) relevant to construction of the mirror a la Greene-Plesser prescription. However, as is common in such calculations (See \cite{DDF,denef_LesHouches,Kachru_et_al}), one assumes that one is working with a subset of periods of $\Gamma$-invariant cycles - the six periods corresponding to the two complex structure deformations in (\ref{eq:hyp_def}) will coincide with the six periods of the mirror - the complex structure moduli absent in (\ref{eq:hyp_def}) will appear only at a higher order in the superpotential because of $\Gamma$-invariance and can be consistently set to zero (See \cite{Kachru_et_al}).

As shown in \cite{D3_D7_Misra_Shukla}, in order to support MSSM (-like) models and for resolving the tension between LVS cosmology and LVS phenemenology within a string theoretic setup, a mobile space-time filling $D3-$brane and stacks of $D7$-branes wrapping the ``big" divisor $\Sigma_B$ along with magnetic fluxes, are included. Working, for concreteness, in the $x_2=1$-coordinate patch and with ther inhomogeneous coordinates: $z_1= \frac{x_1}{x_2},\ z_2=\frac{x_3}{x_2},\ z_3= \frac{x_4}{x_2^6}$ and $z_4=\frac{x_5}{x_2^9}$, the appropriate ${\cal N}=1$ coordinates in the presence of a single $D3$-brane and a single $D7$-brane wrapping the big divisor $\Sigma^B$ along with $D7$-brane fluxes are given as \cite{jockersetal}:
\begin{eqnarray}
\label{eq:N=1_coords}
& & \hskip-0.4cm S = \tau + \kappa_4^2\mu_7{\cal L}_{A{\bar B}}\zeta^A{\bar\zeta}^{\bar B}, \tau=l+ie^{-\phi} \ \ {\cal G}^a = c^a - \tau {\cal B}^a\nonumber\\
& & \hskip-0.4cm T_\alpha=\frac{3i}{2}(\rho_\alpha - \frac{1}{2}\kappa_{\alpha bc}c^b{\cal B}^c) + \frac{3}{4}\kappa_\alpha + \frac{3i}{4(\tau - {\bar\tau})}\kappa_{\alpha bc}{\cal G}^b({\cal G}^c
- {\bar {\cal G}}^c) \nonumber\\
& & \hskip-0.4cm + 3i\kappa_4^2\mu_7l^2C_\alpha^{I{\bar J}}a_I{\bar a_{\bar J}} + \frac{3i}{4}\delta^B_\alpha\tau Q_{\tilde{f}} + \frac{3i}{2}\mu_3l^2(\omega_\alpha)_{i{\bar j}} z^i\bigl({\bar z}^{\bar j}-\frac{i}{2}{\bar z}^{\tilde{a}}({\bar{\cal P}}_{\tilde{a}})^{\bar j}_lz^l\bigr)
\end{eqnarray}
where the axion-dilaton modulus is shifted by a geometric contribution coming from D7-brane moduli and the complexified divisor volume has three new contributions (the second line in expression of $T_\alpha$) namely coming from Wilson line moduli, internal fluxes turned on two cycles and spacetime filling mobile D3-brane fluctuations. The various terms and symbols used above are:
\begin{itemize}
\item
 $\kappa_4$ is related to four-dimensional Newton's constant, $\mu_3$ and $\mu_7$ are $D3$ and $D7$-brane tensions, $\kappa_{\alpha ab}$'s are triple intersection integers of the CY orientifold, and $c^a$ and $b^a$ are coefficients of RR and NS-NS two forms expanded in odd basis of $H^{(1,1)}_{{\bar\partial},-}(CY)$,
\item
 $ {\cal L}_{A{\bar B}}=\frac{\int_{\Sigma^B}\tilde{s}_A\wedge\tilde{s}_{\bar B}}{\int_{CY_3}\Omega\wedge{\bar\Omega}}$,
$\tilde{s}_A\in H^{(2,0)}_{{\bar\partial},-}(\Sigma^B)$,
\item
fluctuations of $D7$-brane in $CY_3$ normal to $\Sigma^B$ are denoted by $\zeta\in H^0(\Sigma^B,N\Sigma^B)$, i.e., they are the space of global sections of the normal bundle $N\Sigma^B$,
\item
${\cal B}\equiv b^a - lf^a$, where $f^a$ are the components of elements of two-form fluxes valued in $i^*\left(H^2_-(CY_3)\right)$, immersion map is defined as:
$i:\Sigma^B\hookrightarrow CY_3$,
\item
 $C^{I{\bar J}}_\alpha=\int_{\Sigma^B}i^*\omega_\alpha\wedge A^I\wedge A^{\bar J}$, $\omega_\alpha\in H^{(1,1)}_{{\bar\partial},+}(CY_3)$ and $A^I$ forming a basis for $H^{(0,1)}_{{\bar\partial},-}(\Sigma^B)$,
\item
 $a_I$ is defined via a Kaluza-Klein reduction of the $U(1)$ gauge field (one-form) $A(x,y)=A_\mu(x)dx^\mu P_-(y)+a_I(x)A^I(y)+{\bar a}_{\bar J}(x){\bar A}^{\bar J}(y)$, where $P_-(y)=1$ if $y\in\Sigma^B$ and -1 if $y\in\sigma(\Sigma^B)$,
\item
 $z^{\tilde{a}}$ are $D=4$ complex structure deformations of the CY orientifold defined via:\\ $\delta g_{{\bar i}{\bar j}}(z^{\tilde{a}})=-\frac{i}{||\Omega||^2}z^{\tilde{a}}\left(\chi_{\tilde{a}}\right)_{{\bar i}jk}\left({\bar\Omega}\right)^{jkl}g_{l{\bar j}}$, where $\left(\chi_{\tilde{a}}\right)_{{\bar i}jk}$ are components of the basis elements of $H^{(2,1)}_{{\bar\partial},-}(CY_3)$,
\item
 $\left({\cal P}_{\tilde{a}}\right)^i_{\bar j}\equiv\frac{1}{||\Omega||^2}{\bar\Omega}^{ikl}\left(\chi_{\tilde{a}}\right)_{kl{\bar j}}$, i.e.,
${\cal P}:TCY_3^{(1,0)}\longrightarrow TCY_3^{(0,1)}$ via the transformation:
$z^i\stackrel{\rm c.s.\ deform}{\longrightarrow}z^i+\frac{i}{2}z^{\tilde{a}}\left({\cal P}_{\tilde{a}}\right)^i_{\bar j}{\bar z}^{\bar j}$,
\item
 $z^i$ are scalar fields corresponding to geometric fluctuations of $D3$-brane inside the Calabi-Yau and defined via: $z(x)=z^i(x)\partial_i + {\bar z}^{\bar i}({\bar x}){\bar\partial}_{\bar i}$, and
\item
 $Q_{\tilde{f}}\equiv l^2\int_{\Sigma^B}\tilde{f}\wedge\tilde{f}$, where $\tilde{f}\in\tilde{H}^2_-(\Sigma^B)\equiv{\rm coker}\left(H^2_-(CY_3)\stackrel{i^*}{\rightarrow}H^2_-(\Sigma^B)\right)$.
\end{itemize}

The involution $\sigma$, as part of the type IIB orientifold ${\cal O}$, is isometric and holomorphic implying the decomposition of $H^{p,q}_{{\bar\partial}}(CY_3)$ into the positive and negative eigen-subspaces of the involution and $\sigma^*J=J$ (See references \cite{jockersetal,Grimm}). For type IIB with space-time filling $D3/D7$-branes, ${\cal O}=(-)^{F_L}\Omega\sigma^*$ with $\sigma^*\Omega=-\Omega$ implying that one will need to include $O3/O7$-planes. Though our subsequent discussion will remain general so that one does have to commit to any particular involution, however one specific choice of involution is considered in appendix {\bf A}, where we also construct a basis for $H^{1,1}_-(CY_3, {\bf Z})$: $dim_{\bf R}H^{1,1}_-(CY_3, {\bf Z}) =2$. The RR-tadpole cancelation conditions for space-time filling $D3$ and $D7$-branes are:
\begin{eqnarray*}
& & \sum_{D7_i}\mu_7\int_{{\bf R}^{1,3}\times\Sigma_B}C^{(8)}+\sum_{O7_i}\nu_7\int_{{\bf R}^{1,3}\times\tilde{\Sigma}_B}C^{(8)}=0,\nonumber\\
 &  & \sum_{D7_i}\mu_7\int_{{\bf R}^{1,3}\times\Sigma_B}C^{(4)}\wedge f_i\wedge f_i
 +\sum_{D3_i}\mu_3\int_{{\bf R}^{1,3}\times\{\Sigma_B\}}C^{(4)}+\sum_{D3_i}\nu_3\int_{{\bf R}^{1,3}\times\{\tilde{\Sigma}_B\}}C^{(4)}=0,
\end{eqnarray*}
where all $D7$-branes wrap $\Sigma_B$ but with different two-form fluxes $f_i$ (turned on two-cycles homologically non-trivial in $\Sigma_B$, and not the Calabi-Yau) and the $O7$-planes wrap the corresponding four-cycle $\tilde{\Sigma}_B$ in the fixed-point locus of $\sigma$; $\mu_7$ and $\nu_7$ are the RR charges of the $D7$-branes and $O7$-planes respectively, and $\{\Sigma_B\}/\{\tilde{\Sigma}_B\}$ denotes the point along $\Sigma_B/\tilde{\Sigma}_B$ specifying the location of the mobile space-time filling $D3$-brane/$O3$-plane.
We do not consider $C^{(6)}$-tadpole cancelation condition as the same, after inclusion of the image branes under the involution, is usually identically satisfied. The NS-NS tadpoles are taken care of either by working  in the dilute flux approximation or by mild supersymmetry breaking wherein the $D$-term potential $\left(\int_{\Sigma_B}i^*J\wedge i^*{\cal F}\right)^2\approx0$ where ${\cal F}\equiv B-2\pi\alpha^\prime f={\cal F}^\alpha {\cal F}^\beta \kappa_{\alpha\beta} + \tilde{\cal F}^a\tilde{\cal F}^b\kappa_{ab}$  where ${\cal F}^\alpha$ and
${\cal F}^a$ are the components of ${\cal F}$ in an expansion in the basis
$i^*\omega_\alpha, \omega_\alpha\in H^{(1,1)}(CY_3)$ and $\tilde{\omega}_a\in coker\left(H^{(1,1)}_-(CY_3)\stackrel{i^*}{\rightarrow}H^{(1,1)}_-(\Sigma_B)\right)$, and $\kappa_{\alpha\beta}=\int_{\Sigma_B}i^*\omega_\alpha\wedge i^*\omega_\beta$ and $\kappa_{ab}=\int_{\Sigma_B}\tilde{\omega}_a\wedge\tilde{\omega}_b$, the immersion map $i$ being defined as:
$i:\Sigma^B\hookrightarrow CY_3$.

The ${\cal N}=1$ orientifold-truncated (single) $D3/D7$-brane spectrum consists of ${\cal N}=1$ vector multiplet $(A_\mu,\lambda)\equiv$ (gauge field, gaugino), three ${\cal N}=1$ chiral multiplets $(z^i,\psi^i), i=1,2,3$, two ${\cal N}=1$ chiral multiplets $(\zeta^A,\chi^A),A=1,..., h^{2,0}_{{\bar\partial},-}(\Sigma_B\cup\sigma(\Sigma_B)), (a_I,\chi_I), I=1,..., h^{0,1}_{{\bar\partial},-}(\Sigma_B\cup\sigma(\Sigma_B))$. The Wilson line moduli $a_I$ and their fermionic superpartners $\chi_I$ play a very important role in our work.

In addition, the orientifold truncation of the bulk massless spectrum is given by the following. Expanding $J, B, C^{(0),(2),(4),(6),(8)}$ as:
\begin{eqnarray*}
& & J=v^\alpha\omega_\alpha, \alpha=1,...,h^{1,1}_+(CY_3),\nonumber\\
& & B=b^a\omega_a; C^{(2)}=C^a\omega_a, a=1,...,h^{1,1}_-(CY_3),\nonumber\\
& & C^{(0)}=l,\nonumber\\
& & C^{(4)}=D^\alpha_{(2)}\wedge\omega_\alpha+V^{\hat{\alpha}}\wedge\alpha_{\hat{\alpha}}+U_{\hat{\alpha}}
\wedge\beta^{\hat{\alpha}}+\rho_\alpha\tilde{\omega}^\alpha, \hat{\alpha}=1,...,h^{2,1}_+(CY_3)\nonumber\\
& & \left[(\alpha_{\hat{\alpha}},\beta^{\hat{\alpha}}):{\rm basis\ for}\ H^{2,1}_{{\bar\partial},+}(CY_3); \tilde{\omega}_\alpha:{\rm basis\ for}\ H^{2,2}_{{\bar\partial},+}(CY_3)\right]\nonumber\\
 & & C^{(6)}=\tilde{c}^2_a\wedge \tilde{\omega}^a, a=1,...,h^{2,2}_{{\bar\partial},-}(CY_3),\nonumber\\
 & & C^{(8)}=\tilde{l}^{(2)}\wedge\frac{\Omega\wedge{\bar\Omega}}{\int_{CY_3}\Omega\wedge{\bar\Omega}},
\end{eqnarray*}
and the D=4 Hodge duality: $$\tilde{l}^{(2)}\sim l, \tilde{c}^2_a\sim C^a, D^\alpha_{(2)}\sim\rho_\alpha, V^{\hat{\alpha}}\sim U_{\hat{\alpha}},$$ one obtains $h^{2,1}_+$ vector multiplets $(V^{\hat{\alpha}})$,
1 chiral multiplet $(l,\phi)$, $h^{1,1}_-$ chiral multiplets $(b^a,c_a)$, $h^{1,1}_+$ chiral multiplets $(\rho_\alpha,v^\alpha)$ and $h^{2,1}_-$ chiral multiplets containing the complex structure deformation scalars ($z^{\tilde{a}=1,...,h^{2,1}_-}$).

In \cite{D3_D7_Misra_Shukla}, it was shown that the potential could be extremized around Higgses' vevs ${\cal V}^{\frac{1}{36}}$ and extremum value of the Wilson line moduli ${\cal V}^{-\frac{1}{4}}$; the fluctuations around the same  are:
\begin{eqnarray}
\label{eq:fluctuations}
& & z_{1,2}={\cal V}^{\frac{1}{36}} + \delta z_{1,2},\nonumber\\
& & {\cal A}_1={\cal V}^{-\frac{1}{4}}+\delta {\cal A}_1.
\end{eqnarray}
The $D7$-brane fluctuation moduli $\zeta_{A(=1,...,h^{(2,0)}_-(\Sigma_B))}$ are set to zero (corresponding to the rigid limit of $\Sigma_B$ and ensures chirality of the spectrum) which also extremizes the ${\cal N}=1$ $F$-term potential.
The soft supersymmetry parameters are related to the expansion of the K\"{a}hler potential and Superpotential for the open- and closed-string moduli as a power series in the open-string (the ``matter fields") moduli.
In terms of the same, the K\"{a}hler potential for the open and closed string moduli, can be written as a power series in the fluctuations $\delta z_i$ and $\delta {\cal A}_I$ as under:
\begin{eqnarray}
\label{eq:K2}
& & \hskip-0.8cm K \left(\left\{\sigma^b,{\bar\sigma}^B;\sigma^S,{\bar\sigma}^S;{\cal G}^a,{\bar{\cal G}}^a;\tau,{\bar\tau}\right\};\left\{z_{1,2},{\bar z}_{1,2};{\cal A}_1,{\bar{\cal A}_1}\right\}\right) = - ln\left(-i(\tau-{\bar\tau})\right) - ln\left(i\int_{CY_3}\Omega\wedge{\bar\Omega}\right)\nonumber\\
 & & \hskip-0.8cm - 2 ln\Biggl[a\left(T_B + {\bar T}_B - \gamma K_{\rm geom}\right)^{\frac{3}{2}}-a\left(T_S + {\bar T}_S - \gamma K_{\rm geom}\right)^{\frac{3}{2}} + \frac{\chi}{2}\sum_{m,n\in{\bf Z}^2/(0,0)}
\frac{({\bar\tau}-\tau)^{\frac{3}{2}}}{(2i)^{\frac{3}{2}}|m+n\tau|^3}\nonumber\\
& & \hskip-0.8cm - 4\sum_{\beta\in H_2^-(CY_3,{\bf Z})} n^0_\beta\sum_{m,n\in{\bf Z}^2/(0,0)}
\frac{({\bar\tau}-\tau)^{\frac{3}{2}}}{(2i)^{\frac{3}{2}}|m+n\tau|^3}cos\left(mk.{\cal B} + nk.c\right)\Biggr]\nonumber\\
 & & \hskip-0.8cm +\frac{C^{KK\ (1)}_s(z^{\tilde{a}},{\bar z}^{\tilde{a}})\sqrt{\tau_s}}{{\cal V}\left(\sum_{(m,n)\in{\bf Z}^2/(0,0)}\frac{\frac{(\tau-{\bar\tau})}{2i}}{|m+n\tau|^2}\right)} + \frac{C^{KK\ (1)}_b(z^{\tilde{a}},{\bar z}^{\tilde{a}})\sqrt{\tau_b}}{{\cal V}\left(\sum_{(m,n)\in{\bf Z}^2/(0,0)}\frac{\frac{(\tau-{\bar\tau})}{2i}}{|m+n\tau|^2}\right)}\nonumber\\
& & \hskip-0.8cm \sim -2 ln\left(\sum_{\beta\in H_2^-(CY_3,{\bf Z})} n^0_\beta(...)\right)  + \left(|\delta z_1|^2 + |\delta z_2|^2 + \delta z_1{\bar\delta z_2} + \delta z_2{\bar\delta z_1}\right)\hat{K}_{z_i{\bar z}_j} + \Biggl\{\left((\delta z_1)^2 + (\delta z_2)^2\right) \nonumber\\
& & \hskip-0.8cm \times \hat{Z}_{z_iz_j} + c.c \Biggr\}+ \Biggl\{|\delta{\cal A}_1|^2\hat{K}_{{\cal A}_1\bar{\cal A}_1} + (\delta {\cal A}_1)^2 \hat{Z}_{{\cal A}_1{\cal A}_1} + c.c \Biggr\}+ \Biggl\{\left(\delta z_1\delta{\bar{\cal A}_1} + \delta z_2\delta{\bar{\cal A}_1} \right)
\hat{K}_{z_i\bar{\cal A}_1} + c.c \Biggr\}\nonumber\\
 & & \hskip-0.8cm +\Biggl\{(\delta z_1\delta{\cal A}_1 + \delta z_2\delta{\cal A}_1 ) \hat{Z}_{z_i{\cal A}_1} +  c.c.\Biggr\} + ....\nonumber\\
\end{eqnarray}
The closed string moduli-dependent K\"{a}hler potential, includes perturbative (using \cite{BBHL}) and non-perturbative (using \cite{Grimm}) $\alpha^\prime$-corrections as well as the loop corrections (using \cite{berghack,loops}). Written out in (discrete subgroup of) $SL(2,{\bf Z})$(expected to survive orientifolding)-covariant form, the perturbative corrections are proportional to $\chi(CY_3)$ and non-perturbative $\alpha^\prime$ corrections are weighted by $\{n^0_\beta\}$, the genus-zero Gopakumar-Vafa invariants that count the number of genus-zero rational curves $\beta\in H_2^-(CY_3,{\bf Z})$. The loop-contributions (dependent also on the complex structure moduli $z^{\tilde{a}}$) arise from KK modes corresponding to closed string or 1-loop open-string exchange between $D3$- and $D7$-(or $O7$-planes)branes wrapped around the ``s" and ``b" divisors. Note,  the two divisors of
${\bf WCP}^4[1,1,1,6,9]$ do not intersect implying that there is no contribution from winding modes corresponding to strings winding non-contractible 1-cycles in the intersection locus corresponding to stacks of intersecting $D7$-branes wrapped around the ``s" and ``b" divisors. One sees from (\ref{eq:K2}) that in the LVS limit, loop corrections are sub-dominant as compared to the perturbative and non-perturbative $\alpha^\prime$ corrections.  In fact, the closed string moduli-dependent contributions are dominated by the genus-zero Gopakumar-Vafa invariants which using Castelnuovo's theory of moduli spaces can be shown to be extremely large for compact projective varieties \cite{Klemm_GV} such as the one used. In (\ref{eq:K2}),
$$\hat{K}_{i{\bar j}}\equiv\frac{\partial^2 K \left(\left\{\sigma^b,{\bar\sigma}^B;\sigma^S,{\bar\sigma}^S;{\cal G}^a,{\bar{\cal G}}^a;\tau,{\bar\tau}\right\};\left\{\delta z_{1,2},{\bar\delta}{\bar z}_{1,2};\delta{\cal A}_1,{\bar\delta}{\bar{\cal A}_1}\right\}\right)}{\partial C^i{\bar\partial} {\bar C}^{\bar j}}|_{C^i=0}$$
and $$\hat{Z}_{ij}\equiv\frac{\partial^2 K \left(\left\{\sigma^b,{\bar\sigma}^B;\sigma^S,{\bar\sigma}^S;{\cal G}^a,{\bar{\cal G}}^a;\tau,{\bar\tau}\right\};\left\{\delta z_{1,2},{\bar\delta}{\bar z}_{1,2};\delta{\cal A}_1,{\bar\delta}{\bar{\cal A}_1}\right\}\right)}{\partial C^i\partial C^j}|_{C^i=0}$$ - the matter field fluctuations denoted by $C^i\equiv \delta z_{1,2},\delta\tilde{\cal A}_1$. As was shown in \cite{D3_D7_Misra_Shukla}, the following basis of fluctuations simultaneously diagonalizes $\hat{K}_{i{\bar j}}$ and $Z_{ij}$:
\begin{eqnarray}
\label{eq:diagonal}
& & \delta\tilde{{\cal A}_1}\equiv (\beta_1\delta z_1 + \beta_2\delta z_2){\cal V}^{-\frac{8}{9}} +
\delta{\cal A}_1,\nonumber\\
& & \delta{\cal Z}_i\equiv \delta z_i + \lambda_2\delta{\cal A}_1{\cal V}^{-\frac{8}{9}}.
\end{eqnarray}
Hence, for evaluating the physical $\mu$ terms, Yukawa couplings, etc., we will work with the fluctuation fields as given in (\ref{eq:diagonal}). Using GLSM techniques and the toric data for the given Swiss-Cheese Calabi-Yau, the geometric K\"{a}hler potential for the divisor ${\Sigma_B}$ (and ${\Sigma_S}$) in the LVS limit was evaluated in \cite{D3_D7_Misra_Shukla} in terms of derivatives of genus-two Siegel theta functions as well as two Fayet-Iliopoulos parameters corresponding to the two $C^*$ actions in the  two-dimensional ${\cal N}=2$ supersymmetric gauge theory whose target space is our toric variety Calabi-Yau, and a parameter $\zeta$ encoding the information about the $D3$-brane position moduli-independent (in the LVS limit) period matrix of the hyperelliptic curve $w^2=P_{\Sigma_B}(z)$, $P_{\Sigma_B}(z)$ being the defining hypersurface for $\Sigma_B$. The geometric K\"{a}hler potential for the divisor $D_5$ in the LVS limit is hence given by:
\begin{eqnarray}
\label{eq:Kaehler_D_5}
& & K|_{D_5} =  r_2 - {\left[r_2 - \left(1+|z_1|^2+|z_2|^2\right)\left(\frac{\zeta}{
r_1|z_3|^2}\right)^{\frac{1}{6}}\right]}\left(\frac{4\sqrt{\zeta}}{3}\right)\nonumber\\
& & +|z_3|^2\left\{{\left[r_2 - \left(1+|z_1|^2+|z_2|^2\right)\left(\frac{\zeta}{
r_1|z_3|^2}\right)^{\frac{1}{6}}\right]}\frac{\sqrt{\zeta}}{3\sqrt{r_1|z_3|^2}}\right\}^2\nonumber\\
& & - r_1 ln\left[\frac{\left(r_2 - \left(1+|z_1|^2+|z_2|^2\right)\left(\frac{\zeta}{
r_1|z_3|^2}\right)^{\frac{1}{6}}\right)\sqrt{\zeta}}{3\sqrt{r_1|z_3|^2}}\right]
-r_2 ln\left[\left(\frac{\zeta}{r_1|z_3|^2}\right)^{\frac{1}{6}}\right]\nonumber\\
& & \sim{{\cal V}^{\frac{2}{3}}}/{\sqrt{ln {\cal V}}}.
\end{eqnarray}
Hence, $\gamma K_{\rm geom}$($\gamma\sim\kappa_4^2\mu_3\sim1/{\cal V}$) and is subdominant as compared to the non-perturbative world-sheet instanton contribution.

Now, in the context of intersecting brane world scenarios \cite{int_brane_SM}, bifundamental leptons and quarks are obtained from open strings stretched between four stacks of $D7$-branes and the adjoint gauge fields correspond to open strings starting and ending on the same $D7$-brane. In Large Volume Scenarios, however,  one considers four stacks of $D7$-branes (the QCD stack of 3 corresponding to $U(3)$, the EW stack of 2 corresponding to $U(2)$ and two single corresponding to $U(1)_Y$ and the EW singlets, just like intersecting brane world models) wrapping $\Sigma_B$ but with different  choices of magnetic $U(1)$ fluxes turned on, on the two-cycles which are non-trivial in the Homology of $\Sigma_B$ and not the ambient Swiss Cheese Calabi-Yau. The inverse gauge coupling constant squared for the $j$-th gauge group ($j:SU(3), SU(2),U(1)$) will be given by $$\frac{1}{g_{j{=SU(3)\ {\rm or}\ SU(2)}}^2} = Re(T_{S/B}) + (F_j^2)\tau,$$
 where
$$F_j^2=F_j^\alpha F_j^\beta \kappa_{\alpha\beta} + \tilde{F}_j^a \tilde{F}_j^b \kappa_{ab},$$ $F_j^\alpha$ are the components of the abelian magnetic fluxes for the $j-$th stack expanded out in the basis of $i^*\omega_\alpha$,  $\omega_\alpha\in H^{1,1}_-(CY_3)$,  and $\tilde{F}_j^a$ are the components of the abelian magnetic fluxes for the the $j$-th stack expanded out in the basis $\tilde{\omega}_a\in coker\left(H^{(1,1)}_-(CY_3)\stackrel{i^*}{\rightarrow}H^{(1,1)}_-(\Sigma_B)\right)$; the intersection matrices $\kappa_{\alpha\beta}=\int_{\Sigma_B}i^*\omega_\alpha\wedge i^*\omega_\beta, \kappa_{ab}=\int_{\Sigma_B}\tilde{\omega}_a\wedge\tilde{\omega}_b$. For ${1}/{g_{U(1)}^2}$ there is a model-dependent numerical prefactor multiplying the right hand side of the $1/g_j^2$-relation. In the dilute flux approximation, $\alpha_i(M_s)/\alpha_i(M_{EW}), i=SU(3),SU(2),U(1)_Y$, are hence unified. By turning on different $U(1)$ fluxes on, e.g., the $3_{QCD}+2_{EW}$ $D7$-brane stacks in the LVS setup, $U(3_{QCD}+2_{EW})$ is broken down to $U(3_{QCD})\times U(2_{EW})$ and the four-dimensional Wilson line moduli $a_{I(=1,...,h^{0,1}_-(\Sigma_B))}$ and their fermionic superpartners $\chi^I$ that are valued, e.g., in the $adj(U(3_{QCD}+2_{EW}))$ to begin with, decompose into the bifundamentals $(3_{QCD},{\bar 2}_{EW})$ and its complex conjugate, corresponding to the bifundamental left-handed quarks of the Standard Model (See \cite{bifund_ferm}).  Further, the main idea then behind realizing $O(1)$ gauge coupling is the competing contribution to the gauge kinetic function (and hence to the gauge coupling) coming from the $D7$-brane Wilson line moduli as compared to the volume of the big divisor $\Sigma_B$, after constructing local (i.e. localized around the location of the mobile $D3$-brane in the Calabi-Yau) appropriate involutively-odd harmonic one-form on the big divisor that lies in $coker\left(H^{(0,1)}_{{\bar\partial},-}(CY_3)\stackrel{i^*}{\rightarrow}
H^{(0,1)}_{{\bar\partial},-}(\Sigma_B)\right)$, the immersion map $i$ being defined as:
$i:\Sigma^B\hookrightarrow CY_3$. This will also entail stabilization of the Wilson line moduli at around $ {\cal V}^{-\frac{1}{4}}$ for vevs of around ${\cal V}^{\frac{1}{36}}$ of the $D3$-brane position moduli, the Higgses in our setup. Extremization of the ${\cal N}=1$ potential, as shown in \cite{D3_D7_Misra_Shukla} and mentioned earlier on, shows that this is indeed true.  This way the gauge couplings corresponding to the gauge theories living on stacks of $D7$ branes wrapping the ``big" divisor $\Sigma_B$ (with different $U(1)$ fluxes on the two-cycles inherited from $\Sigma_B$) will be given by:
$g_{YM}^{-2}=Re(T_B)\sim \mu_3{\cal V}^{\frac{1}{18}}$, $T_B$ being the appropriate ${\cal N}=1$ K\"{a}hler coordinate (summarized in (\ref{eq:N=1_coords}) and the relevant text below the same) and $\mu_3$ related to the $D3$-brane tension,
implying a finite (${\cal O}(1)$) $g_{YM}$ for ${\cal V}\sim10^6$.

Working in the $x_2$=1-coordinate patch, for definiteness, we define inhomogeneous coordinates $z_1={x_1}/{x_2},\ z_2={x_3}/{x_2},\ z_3=
{x_4}/{x_2^6}$ and $z_4={x_5}/{x_2^9}$ three of which get identified with the mobile $D3$-brane position moduli. As explained in \cite{DDF}, a complete set of divisors lying within the K\"{a}hler cone, need be considered so that the complex-structure moduli-dependent superpotential $W_{cs}\sim W_{ED3}$ - the ED3-instanton superpotential - therefore only ${\cal O}(1)$ $D3$-instanton numbers, denoted by $n^s$ corresponding to wrapping of the $ED3$-brane around the small divisor $\Sigma_S$, contribute. We would hence consider either $W_{cs}\sim W_{ED3}(n^s=1)$ for $W\sim W_{ED3}(n^s=1)$ or $W_{cs}=-W_{ED3}(n^s=1)$ with $W\sim W_{ED3}(n^s=2)$.  Hence, the
superpotential in the presence of an $ED3-$instanton is of the type (See \cite{Grimm,Ganor1_2,Maldaetal_Wnp_pref})
\begin{equation}
\label{eq:W_np}
\hskip-0.3in W \sim \left(1 + z_1^{18} + z_2^{18} + z_3^2 - 3\phi_0z_1^6z_2^6\right)^{n_s}\Theta_{n^s}(\tau,{\cal G}^a)e^{in^sT_s},
\end{equation}
where $n^s=1\ {\rm or}\ 2$, is the $D3$-instanton number, the holomorphic pre-factor
$\bigl(1 + z_1^{18} + z_2^{18} + z_3^2$ $ - 3\phi_0z_1^6z_2^6\bigr)^{n_s}$ represents a one-loop
determinant of fluctuations around the $ED3$-instanton due to the modification of the warped volume of $\Sigma_S$ because of the presence of the mobile $D3$-brane and $\Theta_{n^s}(\tau,{\cal G}^a)$
is the holomorphic Jacobi theta function of index $n^s$
defined via $\Theta_{n^s}(\tau,{\cal G}^a)=\sum_{m_a}e^{\frac{i\tau m^2}{2}}e^{in^s {\cal G}^am_a}$
(which encodes the contribution of $D1$-instantons in
an $SL(2,{\bf Z})$-covariant form)where ${\cal G}^a=c^a-\tau{\cal B}^a$ with
${\cal B}^a\equiv b^a - lf^a$,  $f^a$ being the components of  two-form fluxes valued in
$i^*\left(H^2_-(CY_3)\right)$. As $e^{in^sT_s}=e^{-n^s{\rm vol}(\Sigma_S)+i...}$ can be thought of as a section of the inverse divisor bundle $n^s[-\Sigma_S]$, the holomorphic prefactor $(1 + z_1^{18} + z_2^{18} + z_3^2 - 3\phi_0z_1^6z_2^6)^{n_s}$ has to be a section of $n^s[\Sigma_S]$ to compensate and the holomorphic prefactor, a section of $n^s[\Sigma_S]$ having no poles, must have zeros of order $n^s$ on a manifold homotopic to $\Sigma_S$  (See \cite{Ganor1_2}).
Here we consider the rigid limit of wrapping of the $D7$-brane around $\Sigma_B$ (to ensure that there is no obstruction to a chiral matter spectrum) which is effected by considering zero sections of the normal bundle $N\Sigma_B$, which does extremize the potential, and hence there is no superpotential generated due to the fluxes on the world volume of the $D7$-brane \cite{jockersetal}. Also the non-perturbative superpotential due to gaugino condensation on a stack of $N$ $D7$-branes wrapping $\Sigma_B$ will be proportional to $\left(1+z_1^{18}+z_2^{18}+z_3^3-3\phi_0z_1^6z_2^6\right)^{\frac{1}{N}}$ (See \cite{Ganor1_2,Maldaetal_Wnp_pref}) , which according to \cite{Ganor1_2}, vanishes as the mobile $D3$-brane touches the wrapped $D7$-brane. To effect this simplification, we will henceforth be restricting the mobile $D3$-brane to the ``big" divisor $\Sigma_B$. It is for this reason that we are justified in considering a single  wrapped $D7$-brane, which anyway can not effect gaugino condensation.

 Note, even though the hierarchy in the divisor volumes (namely vol$(\Sigma_S)\sim ln {\cal V}$ and vol$(\Sigma_B)\sim {\cal V}^{\frac{2}{3}}$) was obtained in \cite{Balaetal2} in a setup without a mobile space-time filling $D3$-brane and $D7$-branes wrapping a divisor and assuming that the superpotential receives an ${\cal O}(1)$ dominant contribution from the complex structure superpotential, one can argue that using the correct choice of K\"{a}hler coordinates $T_S$ and $T_B$, the divisor volumes stabilize at
 vol$(\Sigma_S)\sim {\cal V}^{\frac{1}{18}}$ and vol$(\Sigma_B)\sim {\cal V}^{\frac{2}{3}}$ - the former
 for ${\cal V}\sim10^6$ is close to $ln{\cal V}$ - in our setup wherein the superpotential receives the dominant contribution from the non-perturbative instanton-generated superpotential. The argument for the same implying stability of large volume compactification or equivalently divisor-volume hierarchy, in our setup was sketched out as a footnote in \cite{Sparticles_Misra_Shukla}; the same is expanded upon in appendix ${\bf B}$.

The superpotential can be expanded about $\langle z_i\rangle\sim {\cal V}^{\frac{1}{36}}$ and extremum vales of the Wilson line moduli ${\cal V}^{-\frac{1}{4}}$ using the basis (\ref{eq:diagonal}) of fluctuations, as:
\begin{eqnarray}
\label{eq:W_exp}
& & \hskip -0.5cm W\sim{\cal V}^{\frac{n^s}{2}}\Theta_{n^s}(\tau,{\cal G}^a)e^{in^sT(\sigma^S,{\bar\sigma^S};{\cal G}^a,{\bar{\cal G}^a};\tau,{\bar\tau})}[1 + (\delta {\cal Z}_1 + \delta {\cal Z}_2)\{n^s{\cal V}^{-\frac{1}{36}} + (in^s\mu_3)^3{\cal V}^{\frac{1}{36}}\}\nonumber\\
& & \hskip -0.5cm +\delta\tilde{{\cal A}}_1\{-[\lambda_1+\lambda_2](in^s\mu_3){\cal V}^{-\frac{31}{36}} - n^s[\lambda_1+\lambda_2]{\cal V}^{-\frac{11}{12}}\}]+ ((\delta {\cal Z}_1)^2 + (\delta {\cal Z}_2)^2)\mu_{{\cal Z}_i{\cal Z}_i}  + \delta {\cal Z}_1\delta {\cal Z}_2\mu_{{\cal Z}_1{\cal Z}_2}\nonumber\\
& & \hskip -0.5cm +(\delta\tilde{{\cal A}}_1)^2\mu_{\tilde{\cal A}_I\tilde{\cal A}_I} + \delta {\cal Z}_1\delta\tilde{{\cal A}}_1\mu_{{\cal Z}_1\tilde{\cal A}_I} + \delta {\cal Z}_2\delta\tilde{{\cal A}}_1\mu_{{\cal Z}_2\tilde{\cal A}_I} + ((\delta {\cal Z}_1)^3 + (\delta {\cal Z}_2)^3)Y_{{\cal Z}_i{\cal Z}_i{\cal Z}_i} + ((\delta {\cal Z}_1)^2\delta {\cal Z}_2 \nonumber\\
& & \hskip -0.5cm + (\delta {\cal Z}_2)^2\delta {\cal Z}_1)Y_{{\cal Z}_i{\cal Z}_i{\cal Z}_j}+ (\delta {\cal Z}_1)^2\delta{\tilde{\cal A}}_1Y_{{\cal Z}_i{\cal Z}_i\tilde{\cal A}_I} + \delta {\cal Z}_1(\delta\tilde{\cal A}_I)^2Y_{{\cal Z}_i\tilde{\cal A}_I\tilde{\cal A}_I} + \delta {\cal Z}_1\delta {\cal Z}_2\delta\tilde{{\cal A}}_1Y_{{\cal Z}_1{\cal Z}_2\tilde{\cal A}_I}\nonumber\\
& & \hskip -0.5cm + (\delta\tilde{\cal A}_I)^3Y_{\tilde{\cal A}_I\tilde{\cal A}_I\tilde{\cal A}_I}+
 ....,
\end{eqnarray}
where $\sigma^S$ ($\sigma^B$) is the volume of the small (big) divisor complexified by the RR four-form axion, the constants $\lambda_{1,2}$ are functions of the extremum values of the closed string moduli and the Fayet-Illiopoulos parameters corresponding to the NLSM (the IR limit of the GLSM) for an underlying ${\cal N}=2$ supersymmetric gauge theory whose target space is the (toric) projective variety we are considering in our work, and the mobile $D3$-brane position moduli fluctuations $\delta{\cal Z}_i$'s and the $D7-$brane Wilson line moduli fluctuations $\delta\tilde{\cal A}_I$'s ($I$ indexes $dim\left(H^{(0,1)}_{{\bar\partial},-}(CY_3)\right)$ and for simplicity we take $I=1$ corresponding to the harmonic one-form referred to earlier on in this section and explicitly constructed in \cite{D3_D7_Misra_Shukla}) are constructed such that they diagonalize the open string moduli metric. Hence, e.g., the relevant Yukawa couplings relevant to the next section, can be evaluated. Here, we also want to emphasise that even though the hierarchy in the divisor volumes (namely vol$(\Sigma_S)\sim ln {\cal V}$ and vol$(\Sigma_B)\sim {\cal V}^{\frac{2}{3}}$) was obtained in \cite{Balaetal2} in a setup without a mobile space-time filling $D3$-brane and $D7$-branes wrapping a divisor and assuming that the superpotential receives an ${\cal O}(1)$ dominant contribution from the complex structure superpotential, one can show that using the correct choice of K\"{a}hler coordinates $T_S$ and $T_B$, the divisor volumes continue to stabilize at the same values in our setup wherein the superpotential receives the dominant contribution from the non-perturbative instanton-generated superpotential \cite{Sparticles_Misra_Shukla}.

As discussed in \cite{Sparticles_Misra_Shukla}, the spacetime filling mobile D3-brane position moduli $z_i$'s and the Wilson line moduli $a_{I}$'s could be respectively identified with Higgses and sparticles (squarks/sleptons) of some (MS)SM like model. In the next section we discuss obtaining the masses of the fermionic superpartners $\chi^I$ of the Wilson line moduli.

\section{Realizing Fermion and Neutrino Mass Scales}

In the context of realizing (MS)SM spectrum in an string theoretic setup, the estimates of the Yukawa couplings are very crucial for several phenomenological aspects- like generating fermion masses, mixing angles etc. and hence study related to the Yukawa coupling computations have been among the focused interesting issues in string phenomenology since long. Usually, ${\cal O}(1)$ Yukawa couplings are natural and hence, to realize the hierarchically suppressed Yukawa couplings in realistic model candidates, is a key-point in understanding the hierarchy of fermion masses and mixing angles. The models of heterotic orbifold \cite{orbifolds1} as well as of intersecting D-branes have been attractive as these lead to suppressed moduli dependent Yukawa couplings \cite{orbifolds2,Yukawas1,Yukawas2} and also non-trivial off-diagonal couplings are allowed. Moreover, in large volume scenarios, interactions (and hence respective couplings) are usually suppressed by string scale instead of Planck scale ($M_s\sim \frac{M_p}{\sqrt{\cal V}}$, where ${\cal V}$ is the Calabi Yau volume) and this makes large volume models more appropriate for generating small Yuakawa couplings- the same being suppressed by some positive rational powers of large Calabi Yau volume. In the context of large volume compactification scenarios, the Yukawa couplings and related studies have been done in local models of magnetized $D7$-branes in the context of IIB string compactifications and it has been observed that the triple overlap of the normalised matter wavefunctions generates the physical Yukawa couplings \cite{bifund_ferm}. In \cite{D3_D7_Misra_Shukla}, in the context of Type IIB large volume Swiss-Cheese $D3/D7$ orientifold compactificaitons, we have computed Yukawa couplings through the power series expansion of the superpotential (and the K\"{a}hler potential) with appropriate normalization and realized a range of values for physical Yukawa couplings.

As we argued in in the previous section, the spacetime filling mobile D3-brane position moduli ${\cal Z}_i$'s and the Wilson line moduli ${\cal A}_{I}$'s could be respectively identified with the two-Higgses and sparticles (squarks and sleptons) of some (MS)SM like model. Now, we look at the fermion sector of the same Wilson line moduli (denoting the fermionic superpartners of ${\cal A}_I$'s as $\chi_I$'s). The fermion bilinear terms in the 4-dimensional effective action, which are generated from $\int d^4 x  e^{\hat{K}/2} \partial_{\alpha}\partial_{\beta} {W} \chi^{\alpha} \chi^{\beta}$, can be given as:\be
\label{eq:fermion_bilinear}
\int d^4 x \left[e^{\hat{K}/2} Y_{\alpha \beta \gamma} {\cal Z}^{\alpha} \chi^{\beta}
\chi^{\gamma} + e^{\hat{K}/2} \frac{{\cal O}_{\alpha \beta \gamma \delta}}{2 M_P}
{\cal Z}^{\alpha} {\cal Z}^{\beta} \chi^{\gamma} \chi^{\delta}\right]
\ee
The first term in the above equation is responsible for generating fermion masses via giving some VEV to Higgs fields while the second term which is a lepton number violating term generates neutrino masses. We will elaborate on these issues in the respective subsections below.

\subsection{Fermion Mass Scales}

The relevant fermionic bilinear terms (the first term in (\ref{eq:fermion_bilinear})) in the four dimensional effective action can be schematically written in terms of canonically normalized superfields ${\cal Z}^i$ and ${\cal A}^I$  as:
$$\int d^4xd^2\theta \,\hat{Y}_{iIJ} {\cal Z}^i {\cal A}^I
{\cal A}^J.$$   Now the fermionic masses are generated through Higgs mechanism by giving VEV to Higgs fields:
$m_{IJ} = {\hat {Y}_{iIJ} <z_i }>$
where $\hat {Y}_{iIJ}$'s are
 ``Physical Yukawas" defined as $$\hat {Y}_{iIJ}= \frac{e^{\hat{K}/2} Y_{iIJ}}
 {\sqrt {K_{i{\bar i}}} \sqrt {K_{I{\bar I}}} \sqrt {K_{J{\bar J}}}}$$ and the Higgs fields
 $z_i$'s are given a vev: $<z_i> \sim {{\cal V}^{\frac{1}{36}}} M_p$
\cite{D3_D7_Misra_Shukla}. 
Next, we discuss the possibility of realizing fermion masses in the range ${\cal O}({\rm MeV-GeV})$ in our setup, possibly corresponding to any of the masses $m_e=0.51$ MeV, $m_u=5$ MeV, $m_d=10$ MeV, $m_s=200$ MeV, $m_c=1.3$ GeV - the first two generation fermion masses \cite{fermionvaluesSM}.


Using (\ref{eq:W_exp}) the physical Standard Model-like ${\cal Z}_i{\cal A}_I^2$ Yukawa couplings  are found to be (as also evaluated in \cite{D3_D7_Misra_Shukla}):
\begin{equation}
\label{eq:Yuks}
\hat{Y}_{{\cal A}_1{\cal A}_1{\cal Z}_i}\sim{\cal V}^{-\frac{199}{72}-\frac{n^s}{2}}.
 \end{equation}
 The reason why our Yukawa couplings differ from those of \cite{Conlon_D3_D7_pheno} are the following.
 To begin with, unlike \cite{Conlon_D3_D7_pheno}, we include non-perturbative world-sheet instanton corrections to K\"{a}ahler
 potential, which can be very large (with appropriate choice of holomorphic isometric involution
  used for orientifolding and the large values of genus-zero Gopakumar-Vafa invariants for compact
  projective varieties) and not always suppressed as assumed in ``LVS" models thus far in the
 literature.
We have a crucial holomorphic pre-factor in superpotential coming form one-loop determinant of instanton fluctuations, which, as explained in section {\bf 2} as well as (in more details in) reference \cite{D3_D7_Misra_Shukla}, has an important impact on SUSY-breaking parameters and couplings, as well as reconciling the requirements of a super-heavy gravitino in the inflationary epoch and a light gravitino in the present era, within the same string theoretic framework.
Further, in our calculations, Wilson line moduli's contribution is incorporated and the same plays a very important role in realizing order one YM couplings and provides a new possibility of supporting (MS)SM on (fluxed) stacks of $D7$-branes wrapping the big divisor unlike other LVS models studied thus far in which the same is supported on stacks of $D7$-branes wrapping the smaller divisor. Also, most of the LVS models studied thus far in the literature, have assumed that the the superpotential will
be receiving the most dominant ${\cal O}(1)$ contribution from the complex-structure
moduli-valued flux superpotential. Based on the discussion of reference \cite{DDF}, as summarized before
equation (7)  of our manuscript, the same is not expected to be true if one considers contribution
to the superpotential from divisors lying within the Kaehler cone.

The leptonic/quark mass, using (\ref{eq:Yuks}), is given by:
${\cal V}^{-\frac{197}{72}-\frac{n^s}{2}}$ in units of $M_p$, which  implies a range of fermion mass $m_{\rm ferm}\sim{\cal O}({\rm MeV-GeV})$ for Calabi Yau volume ${\cal V}\sim {\cal O}(6\times10^5-10^6)$. For example, a mass of $0.5$ MeV could be realized with Calabi Yau volume ${\cal V}\sim 6.2\times 10^5, n^s=2$. In MSSM/2HDM models, up to one loop, the leptonic (quark) masses do not change (appreciably) under an RG flow from the intermediate string scale down to the EW scale (See \cite{Das_Parida}). This way, we show the possibility of realizing all fermion masses of first two generations in our setup. Although we do not have sufficient field content to identify all first two families' fermions, we believe that the same could be realized after inclusion of more Wilson line moduli in the setup. The above results also make the possible identification of Wilson line moduli with squarks and sleptons of first two families  \cite{Sparticles_Misra_Shukla}, more robust.

\subsection{Neutrino Mass scale(s)}

The non-zero Majorana neutrino masses are generated through the Weinberg-type dimension-five operators (the second term in (\ref{eq:fermion_bilinear})) arising from a lepton number violating term written out schematically in terms of canonically normalized superfields ${\cal Z}^i$ and ${\cal A}^I$ as: $$\int d^4x\int d^2\theta e^{\hat{K}/2}\times\left({\cal Z}^2{\cal A}_I^2\in\frac{\partial^2W}{\partial Z^2}{\cal A}_I^2\right)$$ [where ${\cal A}_I=a_I+\theta \chi^I+...$]yielding:
\begin{equation}
\label{eq:nu mass}
m_{\nu}={v^2 sin^2\beta \hat{{\cal O}}_{{\cal Z}_i{\cal Z}_j{\cal Z}_k{\cal Z}_l}}/{2M_p},
\end{equation}
 where $\hat{{\cal O}}_{{\cal Z}_i{\cal Z}_i{\cal Z}_i{\cal Z}_i}$ is the coefficient of the physical/noramalized term quartic in the $D3-$brane position moduli ${\cal Z}_i$ which are defined in terms of diagonal basis of K\"{a}hler potential in ${\cal Z}_i$'s and is given as
$$\hat{{\cal O}}_{{\cal Z}_i{\cal Z}_i{\cal Z}_i{\cal Z}_i}={\frac{{e^\frac{\hat{K}}{2}}{\cal O}_{{\cal Z}_i{\cal Z}_j{\cal Z}_k{\cal Z}_l}}{{\sqrt{\hat{K}_{{\cal Z}_i{\bar{\cal Z}}_{\bar i}}\hat{K}_{{\cal Z}_j{\bar{\cal Z}}_{\bar j}}\hat{K}_{{\cal Z}_k{\bar{\cal Z}}_{\bar k}}\hat{K}_{{\cal Z}_l{\bar{\cal Z}}_{\bar l}}}}}},$$ and
$v sin\beta$ is the vev of the $u$-type Higgs $H_u$ with $sin\beta$ defined via
$tan\beta={\langle H_u\rangle}/{\langle H_d\rangle}$.

Now expanding out superpotential (\ref{eq:W_np}) as a power series in ${\cal Z}_i$, one can show that the coefficient of unnormalized quartic term comes out to be:
$$\hskip-0.1in{\cal O}_{{\cal Z}_i{\cal Z}_j{\cal Z}_k{\cal Z}_l}\sim \frac{2^{n^s}}{24}10^2(\mu_3 n^s l^2)^4{\cal V}^{\frac{n^s}{2}+\frac{1}{9}}
e^{-n^s {\rm vol}(\Sigma_S) + i n^s \mu_3l^2{\cal V}^{\frac{1}{18}}(\alpha+i\beta)}$$
where $\alpha,\beta\sim{\cal O}(1)$ constants and $l=2\pi\alpha^\prime$. Now (as implicitly assumed in
\cite{D3_D7_Misra_Shukla}), $z_i=\gamma_i{\cal V}^{\frac{1}{36}}, i=1,2$ and ${\rm vol}(\Sigma_S)=\gamma_3
ln {\cal V}$ such that $\gamma_3 ln{\cal V} + \mu_3l^2\beta {\cal V}^{\frac{1}{18}} = ln {\cal V}$. Now,
$\mu_3=\frac{1}{(2\pi)^3(\alpha^\prime)^2}$ and hence for $n^s=2$, one obtains $n^s\mu_3l^2=\frac{1}{\pi}$. From \cite{D3_D7_Misra_Shukla}, one uses that $$\hat{K}_{{\cal Z}_i{\bar{\cal Z}}_{\bar i}}\sim\frac{{\cal V}^{\frac{1}{72}}}{\sqrt{\sum_\beta n^0_\beta}},$$ which assuming a holomorphic isometric involution $\sigma$ (as part of the Swiss-Cheese orientifolding action $(-)^{F_L}\Omega\cdot \sigma$) such that $\sum_\beta n^0_\beta\sim\frac{{\cal V}}{10}$, yields $\hat{K}_{{\cal Z}_i{\bar{\cal Z}}_{\bar i}}\sim\sqrt{10}{\cal V}^{-\frac{35}{72}}$. Now, we will elaborate on running of the neutrino mass. We will do so by first discussing the RG flow of $\langle H_u\rangle$ and then the coefficient ($\kappa_{ij}$) of dimension-five operator.

The MSSM RG flow equation for the coefficient $\kappa_{ij}$ of dimension-five operator $\kappa_{ij}L_iH.L_jH$ (responsible for generating neutrino masses) up to one loop were worked out in \cite{Babu_et_al} and are given as under:
\begin{eqnarray}
\label{eq:RG_k_MSSM}
& & 8\pi^2\frac{d\kappa}{d(ln\mu)}=\left(tr\left(3Y^u Y^u\ ^\dagger\right)-4\pi\left(3\alpha_2+\frac{3}{5}\alpha_1\right)\right)\kappa
 + \frac{1}{2}\left(\left(Y^e Y^e\ ^\dagger\right)\kappa + \kappa\left(Y^e Y^e\ ^\dagger\right)^T\right),\nonumber\\
\end{eqnarray}
$Y^u$ and $Y^e$ denoting the up quark and electron Yukawa coupling matrices, and $\alpha_1$ and
$\alpha_2$ denoting the $U(1)$ and $SU(2)$ fine structure constants.
Now, as shown in \cite{Babu_et_al}, unlike MSSM, there are multiple dimension-five operators
in 2HDM corresponding to the Higgses. In our setup, we have taken the two Higgses to be on the same footing
and hence along this locus, there is only one dimension-five operator in the 2HDM as well. In
this (and the LVS) limit(s), and assuming that the Higgses couple only to the $u$-type quarks
as well as taking the $U(1)_Y$ fine structure constant to be equal to the coefficient of the term quartic in the Higgs\footnote{Given that $\lambda\sim(n^s\mu_3l^2)^4\sim1/\pi^4$ in our setup, this would imply, e.g., at the string scale $g_{U(1)}^2\sim0.01$, which is quite reasonable. See also \cite{Sasaki_Carena_Wag}
 for justification.}, one then sees that the 2HDM and MSSM RG flow equations for $\kappa$ become identical. Using then the one-loop solution to the $\langle H_u\rangle$ RG flow equation for the 2HDM as given in \cite{Das_Parida}, one obtains:
\begin{eqnarray}
\label{eq:sol_2HDM_Hu_I}
& &\hskip-0.3in (v sin\beta)_{M_s}=\langle H_u\rangle_{M_s} =\langle H_u\rangle_{M_{EW}}e^{-\frac{3}{16\pi^2}\int_{ln(M_{EW})}^{ln(M_S)}Y^2_t dt^\prime}\biggl(\frac{\alpha_1(M_s)}{\alpha_1(M_{EW})}\biggr)^{3/56}\biggl(\frac{\alpha_2(M_s)}{\alpha_2(M_{EW})}\biggr)^{-3/8},\nonumber\\
\end{eqnarray}
where using arguments of the next paragraph, one can set the exponential to unity.
Now, in the dilute flux approximation, the coupling-constants-dependent factor is
$\left(\frac{\alpha_2(M_s)}{\alpha_2(M_{EW})}\right)^{-9/28}$. Further, using the one-loop RG flow solution for
$\alpha_2$ of \cite{Sparticles_Misra_Shukla}:
\begin{equation}
\label{eq:sol_2HDM_Hu_II}
(v sin\beta)_{M_{EW}}\sim(v sin\beta)_{M_s}\left(1-\frac{{\cal O}(60)}{\frac{16\pi^2}{g^2(M_s)}}\right)^{\frac{9}{28}},
\end{equation}
where $(v sin\beta)_{M_{EW}}\sim246 GeV$ in the large $tan\beta$ regime. Hence, by requiring
$g^2(M_s)$ to be sufficiently close to $16\pi^2/{\cal O}(60)\sim2.5$, one can RG flow $\langle H_u\rangle_{M_s}$ to the required value $\langle H_u\rangle_{M_{EW}}\sim 246 GeV$.

The analytic solution to RG running equation for $\kappa$ is given by:
$$\kappa(M_s)
=\frac{\kappa_{\tau\tau}(M_s)}{\kappa_{\tau\tau}(M_{EW})}\left(\begin{array}{ccc}
\frac{I_e}{I_\tau}& 0 & 0\\
0 & \frac{I_\mu}{I_\tau} & 0 \\
0 & 0 & 1
\end{array}\right)\kappa(M_{EW})\left(\begin{array}{ccc}
\frac{I_e}{I_\tau}& 0 & 0\\
0 & \frac{I_\mu}{I_\tau} & 0 \\
0 & 0 & 1
\end{array}\right)$$
where $I_{e/\mu/\tau}=e^{\frac{1}{8\pi^2}\int_{ln M_{EW}}^{ln M_S}dt\hat{Y}_{e/\mu/\tau}^2}$. Now, for $tan\beta=\frac{\langle z_1\rangle}{\langle z_2\rangle}<50$,
$\frac{I_{e/\mu}}{I_\tau}\approx \left(1-\frac{\hat{Y}_\tau^2}{8\pi^2}ln\left(\frac{M_S}{M_{EW}}\right)\right)$
\cite{RG_neutrino_I}. Assuming $\hat{Y}_t^2/(4\pi)^2\sim10^{-5}$ - see \cite{Ibanez_et_al} - one sees that $I_{e/\mu/\tau}\approx 1$. As $\kappa_{\tau\tau}$ is an overall factor in $\tau$, one can argue that it can be taken to be scale-independent \cite{energy_MNS}. Hence, in MSSM and 2HDM, the coefficient of quartic term $\hat{{\cal O}}_{{\cal Z}_i{\cal Z}_i{\cal Z}_i{\cal Z}_i}$ does not run. For ${\cal V}\sim10^6$ (in string length units), $n^s=2$  and reduced Planck mass $M_p=2.4\times10^{18}{\rm GeV}$,  one obtains:
\begin{equation}
 \label{eq:final_I}
 m_\nu=\frac{(v sin\beta)_{M_{EW}}^2\hat{O}_{{\cal Z}_i{\cal Z}_j{\cal Z}_k{\cal Z}_l}}{2M_p}\stackrel{<}{\sim} 1eV.
 \end{equation}

\section{Proton Decay}

The possibility of proton decay in Grand unified theories is caused by higher dimensional B-number-violating operators. In SUSY and SUGRA GUTs the most important contributions for proton decay come from dimension-four and dimension-five B-number-violating operators (which are model dependent). However for non-supersymmetric GUTs, dimension-six operators are most important (see \cite{Prot_Decay_review} and references therein). Further, in the supersymmetric GUT-like theories, the contributions for proton decay coming from dimension-four operators are usually absent due to symmetries of model and the next crucial and potentially dangerous contributions are due to dimension-five and dimension-six operators. Also gauge dimension-six operators conserve $B-L$ and hence possible decay channels coming from these contributions are a meson and an antilepton, (e.g. $p\longrightarrow K^{+}{\bar\nu}$, $p\longrightarrow \pi^{+}{\bar\nu}$, $p\longrightarrow K^{0}{\bar e}$, $p\longrightarrow \pi^{0}{\bar e}$ etc. \cite{Prot_Decay_review}). Further the B-number-violating dimension-five operators in such (SUSY GUT-type) models relevant to proton decay are of the type: $({\rm squark})^2({\rm quark})({\rm lepton})$ or $({\rm squark})^2({\rm quark})^2$ (See \cite{Prot_Decay_review,EllisNanopRudaz}). This would correspond to ${\partial^2W}/{\partial{\cal A}_I^2}|_{\theta=0}(\chi^I)^2$, in our setup.  From (\ref{eq:W_np}), we see that as long as the mobile $D3-$brane is restricted to $\Sigma_B$, there is no ${\cal A}_I$-dependence of $W$ implying the stability of the proton up to dimension-five operators.

Now, using the notations and technique of \cite{D3_D7_Misra_Shukla}, consider a holomorphic one-form $$A_2=\omega_2(z_1,z_2)dz_1+\tilde{\omega}_2(z_1,z_2)dz_2$$ where $\omega_2(-z_1,z_2)=\omega_2(z_1,z_2), \tilde{\omega}_2(-z_1,z_2)=-\tilde{\omega}_2(z_1,z_2)$ (under $z_1\rightarrow-z_1,z_{2,3}\rightarrow z_{2,3}$) and $$\partial A_2=(1+z_1^{18}+z_2^{18}+z_3^3-\phi_0z_1^6z_2^6)dz_1\wedge dz_2$$ (implying $dA_2|_{\Sigma_B}=0$).
Assuming $\partial_1\tilde{\omega}_2=-\partial_2\omega_2$, then
around $|z_3|\sim{\cal V}^{1/6}, |z_{1,2}|\sim{\cal V}^{1/36}$ - localized around the mobile $D3$-brane - one estimates $$\tilde{\omega}_2(z_1,z_2)\sim z_1^{19}/19+z_2^{18}z_1+\sqrt{\cal V}z_1-\phi_0/7z_1^7z_2^6$$ with $\omega_2(z_1,z_2)=-\tilde{\omega}_2(z_2,z_1)$ in the LVS limit, and utilizing the result of \cite{D3_D7_Misra_Shukla} pertaining to the $I=J=1$-term, one hence obtains: $$i\kappa_4^2\mu_7C_{I{\bar J}}a_I{\bar a}_{\bar J}\sim{\cal V}^{7/6}|a_1|^2+{\cal V}^{2/3}(a_1{\bar a}_{\bar 2}+c.c.)+{\cal V}^{1/6}|a_2|^2,$$ $a_2$ being another Wilson line modulus. The Wilson line moduli $a_I$ can be stabilized at around ${\cal V}^{-1/4}$ (See \cite{D3_D7_Misra_Shukla}) and hence a partial cancelation between $vol(\Sigma_B)$ and
$i\kappa_4^2\mu_7C_{1{\bar 1}}|a_1|^2$ in $T_B$ is possible. Consider fluctuation in $a_2$ about ${\cal V}^{-1/4}$: $a_2\rightarrow{\cal V}^{-1/4}+a_2$. The K\"{a}hler potential, in the LVS limit will then be of the form
\begin{eqnarray*}
& & K\sim-2 ln\biggl[({\cal V}^{1/6}+{\cal V}^{5/12}({\cal A}_2+c.c.)+{\cal V}^{1/6}{\cal A}_2^\dagger{\cal A}_2)^{3/2} +\sum n^0_\beta(...)\biggr]\end{eqnarray*}
 - $a_2$ promoted to the Wilson line modulus superfield ${\cal A}_2$. When expanded in powers of the canonically normalized $\hat{\cal A}_2$, the SUSY GUT-type four-fermion dimension-six proton decay operator obtained from $$\int d^2\theta d^2{\bar\theta}({\cal A}_2)^2
({\cal A}_2^\dagger)^2/M_p^2(\in K(\hat{\cal A}_I,\hat{\cal A}_I^\dagger,...))$$ will yield
$$\frac{\left({\cal V}^{5/4}/\sum n^0_\beta\right)\left(\chi_2^4/M_p^2\right)}{\left(\sqrt{\hat{K}_{{\cal A}_2{\bar{\cal A}}_2}}\right)^4}.$$ Like the single Wilson-line modulus case of \cite{D3_D7_Misra_Shukla}, 
 $$\sqrt{\hat{K}_{{\cal A}_2{\bar{\cal A}}_2}}\sim\frac{{\cal V}^{65/72}}{\sqrt{\sum_\beta n^0_\beta}}.$$ For ${\cal V}\sim 10^6, \sum n^0_\beta\sim{\cal V}/10$(as in the previous section), the numerical factor approximates to $(10^{-9/2}/M_p)^2$. Using arguments of \cite{Prot_Decay_review} and \cite{Witten_Klebanov_proton_decay}, one expects the proton lifetime to be estimated at: $$\frac{{\cal O}(1)\times L_{\Sigma_B}^{-4/3}(10^{9/2}M_p)^4}{(\alpha^2(M_s)m_p^5)},$$ where $L_{\Sigma_B}$ is the Ray-Singer torsion of $\Sigma_B$. $L_{\Sigma_B}$ can in principle be calculated generalizing the large-volume limit of the metric of $\Sigma_B$ worked out in \cite{D3_D7_Misra_Shukla} using GLSM techniques, via the Donaldson algorithm (See \cite{Braun_et_al}). For this letter we assume it to be ${\cal O}(1)$ and obtain an upper bound on the proton lifetime to be around $10^{61}$ years, in conformity with the very large sparticle masses in our setup.

\section{Conclusion}

In the context of type IIB $D3/D7$ Swiss-Cheese orientifold compactification, we showed the possibility of realizing leptonic/quark mass scales along with neutrino mass scales of first two generations,  for Calabi Yau volume ${\cal V}\sim(10^5-10^6)$. We obtain a split SUSY-like scenario (See \cite{DM} for a strict LVS split SUSY scenario) with light fermions and very heavy bosonic superpartners. In the context of proton decay, we argue the absence of SUSY GUT-type dimension-five operators and estimate an upper bound on the proton lifetime to be around $10^{61}$ years from a SUSY GUT-type dimension-six operator.

For getting closer to realizing the Standard Model in our setup - in particular obtaining the three generations of leptons and quarks - we need to do two things. First, as accomplished in section {\bf 4} in the context of construction of dimension-six four-fermion operators contributing to proton decay operators, we need to construct (more than one as done in \cite{D3_D7_Misra_Shukla} and used in this paper till section {\bf 3}) appropriate involutively-odd harmonic one-forms on the big divisor that lie in \hfill\\ $coker\left(H^{(0,1)}_{{\bar\partial},-}(CY_3)\stackrel{i^*}{\rightarrow}
H^{(0,1)}_{{\bar\partial},-}(\Sigma_B)\right)$; the fermionic superpartners of the Wilson line moduli these one-forms couple to, as well as possibly the fermionic superpartners of the sections of the normal bundle to the big divisor, are expected to fill out the fermionic sector of the Standard Model spectrum.   Second, as explained in (the paragraph preceding the one containing equation (\ref{eq:W_np}) of) section {\bf 2}, by considering appropriate number of multiple stacks of $D7$-branes with different choices of $U(1)$ fluxes turned on, on two-cycles non-trivial from the point of homology of the big divisor (as opposed to the Swiss-Cheese Calabi-Yau three-fold), one can obtain bifundamental fermions.

As an aside, guided by our previous results on the geometric K\"{a}hler potential for the big divisor and now using the Donaldson's algorithm, we also estimate the geometric K\"{a}hler potential for the Swiss-Cheese Calabi-Yau (for simplicity, close to the big divisor) used that becomes Ricci-flat in the large volume limit. Some interesting aspects related to hidden- and visible-sector moduli couplings and their cosmological implications, and finite temperature corrections in the $D3/D7$ setup have been looked at in \cite{Pramod_self}.

\section*{Acknowledgement}

PS is supported by a CSIR Senior Research Fellowship. AM would like to thank the high energy theory groups at Imperial College, EFI at the U.Chicago, the MCFP/CSPT at the U.Maryland, NBI at Copenhagan, MPI for Physics at Munich and the MPI for Gravitation at Golm, where part of this work was done and preliminary versions of which were presented as parts of seminars at these institutes. He would also like to thank D.L\"{u}st for helping us become more clear about our setup through his questions at the aforementioned seminar at MPI, Munich, as well as B.Mukhopadhyaya, D.P.Roy/P.N.Pandita and P.Roy for useful clarifications.

\appendix
\section{$H^{1,1}_-(CY_3, {\bf Z})$ and Large Volume Ricci-Flat Swiss Cheese Metrics}
\setcounter{equation}{0} \seceqaa

In this appendix, we show the possible construction of a basis for $H^{1,1}_-(CY_3, {\bf Z})$ with real dimensionality 2. For this, we start with a choice of involution $$\sigma:z_1\rightarrow-z_1, \, \,  z_{2,3}\rightarrow z_{2,3} \, \, ,$$ and consider $B_{i{\bar j}}dz^i\wedge d{\bar z}^{{\bar j}}$ such that only $B_{1{\bar 2}}$ and $B_{1{\bar 3}}$ non-zero. Reality of $B$ implies: $B_{2{\bar 1}}=-\overline{B_{1{\bar 2}}},\ B_{1{\bar 2}}=-\overline{B_{2{\bar 1}}}; B_{3{\bar 1}}=-\overline{B_{1{\bar 3}}},\ B_{1{\bar 3}}=-\overline{B_{3{\bar 1}}}$, which assuming $B_{i{\bar j}}\in{\bf R}$ implies $B_{1{\bar 2}}=-B_{2{\bar 1}}, B_{1{\bar 3}}=-B_{3{\bar 1}}$. Now as basis elements of $H^{1,1}_-(CY_3, {\bf Z})$, we consider
$$\{\omega_1^{-},\omega_2^{-}\}=\{(dz^1\wedge d{\bar z}^{\bar 2} - dz^2\wedge d{\bar z}^{\bar 1}),(dz^1\wedge d{\bar z}^{\bar 3} - dz^3\wedge d{\bar z}^{\bar 1})\}.$$ If the $\omega^a_-$s form a basis for
$H^{(1,1)}_-(CY_3,{\bf Z})$ - a real subspace of  $H^{1,1}(CY_3)$  then:
$$\int_{CY_3}\omega_a^-\wedge *_6\overline{\omega_b^-}=\int_{CY_3}\omega_a^-\wedge \tilde{\omega}^b_-=\delta^b_a,$$ where $\tilde{\omega}^a_-$'s form a basis for $H^{2,2}_-(CY_3,{\bf Z})$. As $\omega_a^-\in{\bf R}$, there is no need to complex conjugate the Hodge dual of the same when taking the inner product of two such (1,1) forms in implementing the completeness requirement for $\omega_a^-$. Note that $*_n:H^{p,q}\rightarrow H^{n-q,n-p}$. Now,
\begin{eqnarray}
\label{eq:6D_complex *}
& & *_n\omega_{i_1...i_p,{\bar j}_1,...,{\bar j}_q}dz^{i_1}\wedge ...\wedge dz^{i_p}\wedge d{\bar z}^{\bar j_1}\wedge ...\wedge d{\bar z}^{\bar j_q}\nonumber\\
& & \sim\sqrt{g}\epsilon^{i_1...i_p}\ _{{\bar i}_{p+1}...{\bar i}_n}
\epsilon^{{\bar j}_1...{\bar j}_q}\ _{j_{q+1}...j_n}\omega_{i_1...i_p,{\bar j}_1,...,{\bar j}_q}
d{\bar z}^{{\bar i}_{p+1}}\wedge...d{\bar z}^{{\bar i}_n}\wedge dz^{j_{q+1}}\wedge...dz^{j_n}.
\end{eqnarray}
Hence, assuming locally a diagonal Calabi-Yau metric,
\begin{eqnarray}
\label{eq:Hodge_duals}
& & \hskip-0.6in*_6\omega_1^-\sim\sqrt{g}\epsilon^1\ _{{\bar 2}{\bar 3}}\epsilon^{\bar 2}\ _{31}d{\bar z}^{\bar 2}\wedge d{\bar z}^{\bar 3}\wedge dz^3\wedge dz^1 - \epsilon^2\ _{{\bar 3}{\bar 1}}\epsilon^{\bar 1}\ _{23}d{\bar z}^{\bar 3}\wedge d{\bar z}^{\bar 1}\wedge dz^2\wedge dz^3,
\end{eqnarray}
implying
\begin{eqnarray}
\label{eq:completeness_1}
& & \hskip-0.8in\omega_1^-\wedge*_6\overline{\omega_1^-}= \omega_1^-\wedge*_6\omega_1^-\sim2\sqrt{g}dz^1\wedge d{\bar z}^{\bar 1}\wedge dz^2\wedge d{\bar z}^{\bar 2}\wedge dz^3\wedge d{\bar z}^{\bar 3}\sim{\rm vol\ form},
\end{eqnarray}
as well as:
\begin{eqnarray}
\label{eq:completeness_2}
& & \omega_2^-\wedge*_6\overline{\omega_1^-}=\omega_2^-\wedge*_6\omega_1^-=0.
\end{eqnarray}
Similarly, one can argue $\omega_2^-\wedge*_6\omega_2^-\sim$ volume-form; $\{\frac{\omega^1_-}{\sqrt{\cal V}},\frac{\omega^2_-}{\sqrt{\cal V}}\}$ hence forms the required basis for $H^{1,1}_-(CY_3,{\bf Z})$. For a more exact calculation, one can show that:
\begin{eqnarray}
\label{eq:eq:completeness_exact 1}
& & \omega_1^-\wedge*_6\omega_1^- \sim2\sqrt{g}\Biggl[\left(g_{2{\bar 3}}g_{3{\bar 1}}-g_{2{\bar 1}}g_{3{\bar 3}}\right)\left(g_{2{\bar 3}}g_{3{\bar 1}}-g_{2{\bar 1}}g_{3{\bar 3}}\right)-\left(g_{2{\bar 2}}g_{3{\bar 3}}-g_{2{\bar 3}}g_{3{\bar 2}}\right)\nonumber\\
& & \times\left(g_{1{\bar 3}}g_{3{\bar 1}}-g_{1{\bar 1}}g_{3{\bar 3}}\right)+c.c.\Biggr]\sqrt{g}dz^1\wedge d{\bar z}^{\bar 1}\wedge dz^2\wedge d{\bar z}^{\bar 2}\wedge dz^3\wedge d{\bar z}^{\bar 3}
\end{eqnarray}
whereas:
\begin{eqnarray}
\label{eq:eq:completeness_exact 1}
& & \omega_2^-\wedge*_6\omega_1^-\sim2\Biggl[\left(g_{2{\bar 1}}g_{3{\bar 2}}-g_{2{\bar 2}}g_{3{\bar 1}}\right)\left(g_{2{\bar 3}}g_{3{\bar 1}}-g_{2{\bar 1}}g_{3{\bar 3}}\right)-\left(g_{2{\bar 2}}g_{3{\bar 3}}-g_{2{\bar 3}}g_{3{\bar 2}}\right)\nonumber\\
& & \times\left(g_{1{\bar 3}}g_{2{\bar 1}}-g_{1{\bar 1}}g_{2{\bar 3}}\right)+c.c.\Biggr]\sqrt{g}dz^1\wedge d{\bar z}^{\bar 1}\wedge dz^2\wedge d{\bar z}^{\bar 2}\wedge dz^3\wedge d{\bar z}^{\bar 3}.
\end{eqnarray}
To get some idea about $g_{i{\bar j}}$, we will look at the LVS limit of the geometric K\"{a}hler potential of $\Sigma_B$ obtained using GLSM techniques. One sees that:
\begin{equation}
\label{eq:metric-Big}
g_{i{\bar j}}|_{\Sigma_B}\sim \left(\begin{array}{cc}
-\frac{\left(\frac{\zeta }{\sqrt{\ln {\cal V}}}\right)^{7/6}}{6 \sqrt[9]{2}
   }{\cal V}^{\frac{2}{9}}-\frac{\left(\frac{\zeta }{\sqrt{\ln {\cal V}}}\right)^{7/6}}{3 \sqrt[9]{2}
   }{\cal V}^{\frac{5}{18}} & -\frac{\left(\frac{\zeta }{\sqrt{\ln {\cal V}}}\right)^{7/6}}{6 \sqrt[9]{2}
   }{\cal V}^{\frac{2}{9}}+\frac{\left(\frac{\zeta }{\sqrt{\ln {\cal V}}}\right)^{7/6}}{3 \sqrt[9]{2}
   }{\cal V}^{\frac{5}{18}}\\
-\frac{\left(\frac{\zeta }{\sqrt{\ln {\cal V}}}\right)^{7/6}}{6 \sqrt[9]{2}
   }{\cal V}^{\frac{2}{9}}+\frac{\left(\frac{\zeta }{\sqrt{\ln {\cal V}}}\right)^{7/6}}{3 \sqrt[9]{2}
   }{\cal V}^{\frac{5}{18}}& -\frac{\left(\frac{\zeta }{\sqrt{\ln {\cal V}}}\right)^{7/6}}{6 \sqrt[9]{2}
   }{\cal V}^{\frac{2}{9}}-\frac{\left(\frac{\zeta }{\sqrt{\ln {\cal V}}}\right)^{7/6}}{3 \sqrt[9]{2}
   }{\cal V}^{\frac{5}{18}}
   \end{array}\right).
\end{equation}
For ${\cal V}\sim10^{5-6}$, $g_{i\neq j}/g_{i{\bar i}}<1$. We expect this to hold for the full Calabi-Yau. Of course, GLSM techniques do not necessarily guarantee a Ricci-flat metric - the understanding is that the metric so obtained can be appropriately modified by using the Donaldson's algorithm\cite{Donaldson_i}. This is done as follows (Details would be given in \cite{DM}). For simplicity, working near $x_5=0$ - setting $x_5=\epsilon$ - the no-where vanishing holomorphic three-form $$\Omega=\oint \frac{dz_1\wedge dz_2\wedge dz_3\wedge dz_4}{P(\{z_i\})},$$ in the $x_2\neq0$-patch with $z_1=\frac{x_1}{x_2}, z_2=\frac{x_3}{x_2}, z_3=\frac{x_4}{x_2^6}, z_4=\frac{x_5}{x_2^6}$
and $$P(\{z_i\})=1+z_1^{18}+z_2^{18}+z_3^3-\psi\prod_{i=1}^4z_i -\phi z_1^6z_2^6.$$ By the Griffiths residue formula, one obtains: $$\Omega=\frac{dz_1\wedge dz_2\wedge dz_4}{\frac{\partial P}{\partial z_3}}=\frac{dz_1\wedge dz_2\wedge dz_4}{3z_3^2-\psi z_1z_2z_3},$$ which near $z_4\sim\epsilon$ gives $$\frac{dz_1\wedge dz_2\wedge dz_4}{3(\phi z_1^6z_2^6-z_1^{18}-z_2^{18}-1)^{\frac{2}{3}}-\psi\epsilon z_1z_2}.$$ The crux of the Donaldson's algorithm is that the sequence $$\frac{1}{k\pi}\partial_i{\bar\partial}_{\bar j}ln\sum_{\alpha,\beta}h^{\alpha{\bar\beta}}s_\alpha{\bar s}_{\bar\beta}$$ on $P(\{z_i\})$, in the $k\rightarrow\infty$-limit - which in practice implies $k\sim10$ - converges to a unique Calabi-Yau metric for the given K\"{a}hler class and complex structure; $h_{\alpha{\bar\beta}}$ is a balanced metric on the line bundle ${\cal O}_{P(\{z_i\})}(k)$ (with sections $s_\alpha$) for any $k\geq1$, i.e., $$T(h)_{\alpha{\bar\beta}}\equiv \frac{N_k}{\sum_{j=1}w_j}\sum_{i}\frac{s_\alpha(p_i)\overline{s_\beta(p_i)}w_i}
{h^{\gamma{\bar\delta}}
s_\gamma(p_i)\overline{s_\delta(p_i)}}=h_{\alpha{\bar\beta}},$$ where the weight at point $p_i$, $w_i\sim\frac{i^*(J_{GLSM}^3)}{\Omega\wedge{\bar\Omega}}$ with the embedding map $i:P(\{z_i\})\hookrightarrow{\bf WCP}^4$ and the number of sections is denoted by $N_k$. The above corresponds to a K\"{a}hler potential $$K=\frac{1}{k\pi} \, \, ln\sum_{\begin{array}{c}
i_1,...,i_k\\
{\bar j}_1,...,{\bar j}_{\bar k}
\end{array}}h^{(i_1...i_k),({\bar j}_{\bar 1}...{\bar j}_{\bar k})}z_{i_1}...z_{i_k}{\bar z}_{{\bar j}_{\bar 1}...{\bar j}_{{\bar k}}}$$ - the argument of the logarithm being of holomorphic, anti-holomorphic bidegree $(k,k)$. For simplicity, consider $k=2$ for which the sections $s_\alpha$ are given by monomials $z_1^{n_1}z_2^{n_2}z_3^{n_3}$ with $n_1+n_2+n_3\leq2$. Based on our earlier estimate of the geometric K\"{a}hler potential for $\Sigma_B$, we take the following ansatz for the geometric K\"{a}hler potential for the $CY_3$:
\begin{eqnarray}
& & K=- r_1 ln\Biggl[\frac{1}{3\sqrt{r_1|z_1^{18}+z_2^{18}-\phi z_1^6z_2^6|^{\frac{2}{3}}}}\Biggl(r_2 - \frac{{\cal V}^{-\frac{1}{18}}}{h^{z_1^2{\bar z}_1^2}}\left(|z_1|^2+|z_2|^2+z_1{\bar z}_2+{\bar z}_1z_2\right)\nonumber\\
& & -\frac{{\cal V}^{-\frac{1}{12}}\epsilon}{h^{z_1^2{\bar z}_1^2}}\left(z_1{\bar z}_4+{\bar z}_1z_4+z_2{\bar z}_4+{\bar z}_2z_4\right)+h^{z_1^2{\bar z}_1^2}\left(|z_1|^4+|z_2|^4+z_1^2{\bar z}_2^2+z_2^2{\bar z}_1^2+|z_1|^2(z_1{\bar z}_2+{\bar z}_1z_2)\right)\nonumber\\
& & +|z_1|^2\left(z_1{\bar z}_2+{\bar z}_1z_2+|z_1|^2|z_2|^2\right)+\frac{{\cal V}^{-\frac{1}{36}}\epsilon}{h^{z_1^2{\bar z}_1^2}}\biggl(z_1^2{\bar z}_2{\bar z}_4+{\bar z}_1^2z_2z_4+z_2^2{\bar z}_1{\bar z}_4+{\bar z}_2^2z_1z_4\nonumber\\
& & +|z_1|^2\left(z_1{\bar z}_4+{\bar z}_1{\bar z}_4\right)
+ |z_2|^2\left(z_2{\bar z}_4+{\bar z}_2{\bar z}_4\right)+|z_1|^2\left(z_2{\bar z}_4+{\bar z}_2{\bar z}_4\right)
+ |z_2|^2\left(z_1{\bar z}_4+{\bar z}_1{\bar z}_4\right)\biggr)\Biggr)\sqrt{\zeta}
\Biggr]\nonumber\\
& & -r_2 ln\left[\left(\frac{\zeta}{r_1|z_1^{18}+z_2^{18}-\phi z_1^6z_2^6|^{\frac{2}{3}}}\right)^{\frac{1}{6}}\right],
\end{eqnarray}
where the balanced-metric and Ricci-flatness conditions are used to determine the unknown $h^{z_1^2,{\bar z}_1^2}$. Now, with $$w_i\sim\frac{g_{z_1{\bar z}_{\bar 1}}g_{z_2{\bar z}_{\bar 2}}g_{z_4{\bar z}_{\bar 4}}}{|3(\phi z_1^6z_2^6-z_1^{18}-z_2^{18}-1)^{\frac{2}{3}}-\psi\epsilon z_1z_2|^2},$$ we will approximate $\frac{N_kw_i}{\sum_jw_j}\sim{\cal O}(1)$ localizing around the position of $D3$-brane, and in obvious notations and around $z_4\sim\epsilon$, the following is utilized in writing out the above ansatz for the K\"{a}hler potential:
 \begin{eqnarray}
 \label{eq:bal_cond}
& & \sum_{\alpha{\bar\beta}}h^{\alpha{\bar\beta}}s_\alpha{\bar s}_{\bar\beta}\sim h^{z_1^2{\bar z}_{\bar 1}^2}z_1^2{\bar z}_{\bar 1}^2\sim h^{z_1^2{\bar z}_{\bar 1}^2}{\cal V}^{\frac{1}{9}},\nonumber\\
& & T(h)_{z_iz_j{\bar z}_{\bar l}{\bar z}_{\bar 4}}\sim\frac{{\cal V}^{\frac{1}{12}}\epsilon}{h^{z_1^2{\bar z}_{\bar 1}^2}{\cal V}^{\frac{1}{9}}}\sim T(h)_{z_iz_4{\bar z}_{\bar k}{\bar z}_{\bar l}}\sim h_{z_iz_4{\bar z}_{\bar k}{\bar z}_{\bar l}}, \nonumber\\
& & T(h)_{z_iz_j{\bar z}_{\bar 4}^2}\sim T(h)_{z_iz_4{\bar z}_{\bar k}{\bar z}_{\bar 4}}
\sim\frac{{\cal V}^{\frac{1}{18}}\epsilon^2}{h^{z_1^2{\bar z}_{\bar 1}^2}{\cal V}^{\frac{1}{9}}}
\sim h_{z_iz_j{\bar z}_{\bar 4}^2}\sim0,\nonumber\\
& & T(h)_{z_iz_4{\bar z}_{\bar 4}^2} \sim\frac{{\cal V}^{\frac{1}{36}}\epsilon^3}{h^{z_1^2{\bar z}_{\bar 1}^2}{\cal V}^{\frac{1}{9}}} \sim h_{z_iz_4{\bar z}_{\bar 4}^2}\sim0,\nonumber\\
& & T(h)_{z_i{\bar z}_{\bar j}}\sim\frac{h^{z_i{\bar z}_{\bar j}}z_i{\bar z}_{\bar j}}{h^{z_1^2{\bar z}_{\bar 1}^2}{\cal V}^{\frac{1}{9}}}\Biggr|_{{\rm setting}\ h^{z_i{\bar z}_{\bar j}}=1}\sim\frac{{\cal V}^{-\frac{1}{18}}}{h^{z_1^2{\bar z}_{\bar 1}^2}}\sim h_{z_i{\bar z}_{\bar j}},\nonumber\\
& & T(h)_{z_i{\bar z}_{\bar 4}}\sim\frac{h^{z_i{\bar z}_{\bar j}}z_i\epsilon}{h^{z_1^2{\bar z}_{\bar 1}^2}{\cal V}^{\frac{1}{9}}}\Biggr|_{{\rm setting}\ h^{z_i{\bar z}_{\bar 4}}=1}\sim
\frac{{\cal V}^{-\frac{1}{12}}\epsilon}{h^{z_1^2{\bar z}_{\bar 1}^2}}\sim h_{z_i{\bar z}_{\bar 4}};
 \end{eqnarray}
 we have dropped terms of ${\cal O}(\epsilon^2)$ in (\ref{eq:bal_cond}). Further, for a K\"{a}hler manifold, utilizing $\Gamma^l_{jk}=g^{l{\bar m}}\partial_j g_{k{\bar m}}$
and hence $R_{i{\bar j}}=-{\bar\partial}_{\bar j}\Gamma^k_{ik}$. One notes that for $z_1\sim {\cal V}^{\frac{1}{36}}$ and $z_2\sim 1.3{\cal V}^{\frac{1}{36}}$ and $r_2\sim{\cal V}^{\frac{1}{3}}$
and $r_1\sim\sqrt{ln {\cal V}}$:
$$R_{z_1{\bar z}_1}\sim R_{z_2{\bar z}_2}\sim R_{z_1{\bar z}_{\bar 2}}\sim\frac{\left[{\cal V}^{\frac{7}{3}}-\left(h^{z_1^2{\bar z}_{\bar 1}^2}\right)^2{\cal V}^{\frac{20}{9}}+{\cal V}^{\frac{20}{9}}\right]}{{\cal V}^{\frac{41}{18}}}+{\cal O}({\cal V}^{-\frac{5}{18}}),$$
implying that the Ricci tensor, in the LVS limit, vanishes  for
 $$h^{z_1^2{\bar z}_{\bar 1}^2}\sim\frac{\sqrt{{\cal V}^{\frac{20}{9}}+{\cal V}^{\frac{7}{3}}}}{{\cal V}^{\frac{10}{9}}}.$$
 Hence, the LVS Ricci-flat metric's components near $(z_1,z_2,z_4)\sim({\cal V}^{\frac{1}{36}},{\cal V}^{\frac{1}{36}},\epsilon)$ are estimated to be:
 \begin{eqnarray}
 \label{eq:Ricci_flat_metric}
 & & g_{i{\bar j}}\sim\left(\begin{array}{ccc}
\sqrt{ln {\cal V}} \, \, \, {\cal V}^{-\frac{2}{9}} &
\sqrt{ln {\cal V}}\, \, \, {\cal V}^{-\frac{2}{9}} &
\biggl(-{\cal O}(1)\sqrt{ln {\cal V}}\, \, {\cal V}^{-\frac{13}{36}}\\
& & +{\cal O}(1)\sqrt{ln {\cal V}}\, \, {\cal V}^{-\frac{17}{36}}\biggr) \\
& & \\
\sqrt{ln {\cal V}}\, \, \, {\cal V}^{-\frac{2}{9}} &  \sqrt{ln {\cal V}}\, \, \, {\cal V}^{-\frac{2}{9}} &
\biggl(-{\cal O}(1)\sqrt{ln {\cal V}}\, \, {\cal V}^{-\frac{13}{36}}\\
& & +{\cal O}(1)\sqrt{ln {\cal V}}\, \, {\cal V}^{-\frac{17}{36}}\biggr)\\
& & \\
\biggl(-{\cal O}(1)\sqrt{ln {\cal V}}\, \, {\cal V}^{-\frac{13}{36}} &
\biggl(-{\cal O}(1)\sqrt{ln {\cal V}}\, \, {\cal V}^{-\frac{13}{36}}&
\epsilon^2\\
+{\cal O}(1)\sqrt{ln {\cal V}}\, \, {\cal V}^{-\frac{17}{36}}\biggr)& +{\cal O}(1)\sqrt{ln {\cal V}}\, \, {\cal V}^{-\frac{17}{36}} \biggr)& \\
\end{array}\right)\nonumber\\
& & \nonumber\\
& & \hskip 0.6cm \sim\left(\begin{array}{ccc}
{\cal V}^{-\frac{2}{9}} & {\cal V}^{-\frac{2}{9}} & 0\\
{\cal V}^{-\frac{2}{9}} & {\cal V}^{-\frac{2}{9}} & 0\\
0 & 0 & \epsilon^2 \end{array}\right).
\end{eqnarray}
Hence, $$\frac{\int_{CY_3}\omega_a^-\wedge *_6\omega_b^-}{\int_{CY_3}\omega_a^-\wedge *_6\omega_a^-}<<1, a\neq b$$ implying that the completeness relation is approximately satisfied (in the LVS limit).

\section{K\"{a}hler Hirarchy in Divisors' Volumes in $D3/D7$ Setup}
\setcounter{equation}{0} \seceqbb

In this appendix, we reason out that even in the $D3/D7$ Swiss-Cheese setup, one can introduce a hierarchy in the divisor volumes of the small and big divisors of the Swiss Cheese Calabi-Yau (\ref{eq:hyp_def}). A quick way to see this is to extremize the potential with respect to $\tau_S\equiv{\rm vol}(\Sigma_S)$ and see at what value the same stabilizes assuming in a self-consistent manner that $\tau_B\equiv{\rm vol}(\Sigma_B)$ has been stabilized at ${\cal V}^{\frac{2}{3}}$. Referring to the discussion in \cite{D3_D7_Misra_Shukla} on the complete K\"{a}hler potential using the $T_B$ and $T_S$ for $D3-D7$ system and {\it the discussion in the paragraph preceding the one containing equation (\ref{eq:W_np})}, one sees that with the D3-brane restricted to $\Sigma_B$ and for vevs of the mobile $D3$-brane position moduli of around $\langle z_i\rangle\sim {\cal V}^{\frac{1}{36}}$ and (as shown in reference \cite{D3_D7_Misra_Shukla}) assuming a partial cancelation between the contribution of vol$(\Sigma_B)$ and the quadratic Wilson line moduli contribution in $T_B$ to yield ${\cal V}^{\frac{1}{18}}$ having stabilized Wilson line moduli at ${\cal V}^{-\frac{1}{4}}$, one sees that with the $D3$-brane restricted to $\Sigma_B$,
\begin{equation}
\label{eq:K}
K=-2 ln[(\mu_3{\cal V}^{\frac{1}{18}}+...)^{\frac{3}{2}} - (\tau_S+\mu_3{\cal V}^{\frac{1}{18}}+...)^{\frac{3}{2}}+\sum_{\beta\in H_2^-}\sum_{m,n\neq(0,0)}n^0_\beta \frac{({\bar\tau}-\tau)^{\frac{3}{2}}}{(2i)^{\frac{3}{2}}|m+n\tau|^3} cos(nk.b+mk.c)+...] + ...,
\end{equation}
the dots denoting the sub-dominant terms and $\sigma_S=\tau_S+i\rho^{(4)}_S$, $\tau_S$=vol$(\Sigma_S)$ and $\rho^{(4)}_S$ being a $C^{(4)}$ axion -  the contribution from the $n^0_\beta$s, the genus-zero Gopakumar-Vafa invariants, is the most dominant. Further, with the mobile $D3$-brane restricted to $\Sigma_B$ and $\langle z_i\rangle\sim {\cal V}^{\frac{1}{36}}$, $W\sim W_{np}\sim {\cal V}^{\frac{n^s}{2}}e^{-n^s\tau_S}$ - see \cite{DDF} and the discussion in the paragraph containing (\ref{eq:W_np}). As argued in \cite{D3_D7_Misra_Shukla}, one hence sees that the ${\cal N}=1$ potential (setting $\alpha^\prime$ to 1) is given by:
\begin{eqnarray}
\label{eq:N=1_V}
& & V\sim e^K K^{\sigma_S{\bar\sigma}_S}|D_{\tau_S}W_{np}|^2\sim
e^{-2n^s\tau_S}\nonumber\\
& &\hskip-1in\times \sum_{m^a:D1-{\rm inst\ no}}e^{-\frac{m^2}{g_s}}e^{\frac{n^s{\cal B}\cdot m}{g_s}-\frac{2n^s\kappa_{Sbc}{\cal B}^b{\cal B}^c}{g_s}}{\cal V}^{\frac{n^s}{2}}\sqrt{\tau_S + \mu_3{\cal V}^{\frac{1}{18}}}\left[\left(\tau_S + \mu_3{\cal V}^{\frac{1}{18}}\right)^{\frac{3}{2}}+\Xi_0\right]\left[-n^s+\frac{\sqrt{\tau_S + \mu_3{\cal V}^{\frac{1}{18}}}}{\left(\tau_s+\mu_3{\cal V}^{\frac{1}{18}}\right)^{\frac{3}{2}}+{\Xi_0}}\right]^2,\nonumber\\
& &
\end{eqnarray}
where $\Xi_0\sim \sum_{\beta\in H_2^-}n^0_\beta cos(nk.{\cal B}_0+mk.c_0)g_s^{\frac{3}{2}}$,  ${\cal B}^a_0$ and $c^a_0$ being the extremum values of the RR and NS-NS two-form axions. For a generic $\Xi$ in (\ref{eq:N=1_V}) and that according to the Castelnuovo's theory of moduli spaces applied to compact projective varieties - see \cite{Klemm_GV} - $n^0_\beta$ are very large, one can show that the extremum values of the $C^{(2)}$ and $B^{(2)}$ axions will be determined by the following:
\begin{equation}
\label{eq:extr axions}
\sum_\beta n^0_\beta k^a sin(mk.{\cal B}+nk.c)g_s^{\frac{3}{2}}\sim\frac{m^a-4\kappa_{Sab}{\cal B}^b}{\sum_\beta n^0_\beta cos(mk.{\cal B}+nk.c)g_s^{\frac{3}{2}} - {\cal O}(1)},
\end{equation}
implying it is possible to self-consistently stay close to the locus: $mk.{\cal B}+nk.c= l\pi, l\in{\bf Z}$. Henceforth we assume that the holomorphic isometric involution is such that $\sum_\beta n^0_\beta cos(mk.{\cal B}+nk.c)g_s^{\frac{3}{2}}\sim {\cal V}$, which given the earlier remark about the large values of the (largest) Gopakumar-Vafa invariant, is justified. From (\ref{eq:N=1_V}), one obtains:
\begin{eqnarray}
\label{eq:dV}
& & V^\prime(\tau_s)\sim e^{-2n^s\tau_s}\Biggl[
\frac{\left\{\left(\tau_s+\mu_3{\cal V}^{\frac{1}{18}}\right)^{\frac{3}{2}}+\Xi_0\right\}}{\sqrt{\tau_s+\mu_3{\cal V}^{\frac{1}{18}}}}\left(-n^s+\frac{\sqrt{\tau_s+\mu_3{\cal V}^{\frac{1}{18}}}}{\left(\tau_s+\mu_3{\cal V}^{\frac{1}{18}}\right)^{\frac{3}{2}}+{\Xi_0}}\right)^2\nonumber\\
& & +\left(\tau_s+{\cal V}^{\frac{1}{18}}\right)\left(-n^s+\frac{\sqrt{\tau_s+\mu_3{\cal V}^{\frac{1}{18}}}}{\left(\tau_s+\mu_3{\cal V}^{\frac{1}{18}}\right)^{\frac{3}{2}}+{\Xi_0}}\right)^2\nonumber\\
& & +\sqrt{\tau_s+\mu_3{\cal V}^{\frac{1}{18}}}\left(\left(\tau_s+\mu_3{\cal V}^{\frac{1}{18}}\right)^{\frac{3}{2}}+{\Xi_0}\right)\left(-n^s+\frac{\sqrt{\tau_s+\mu_3{\cal V}^{\frac{1}{18}}}}{\left(\tau_s+\mu_3{\cal V}^{\frac{1}{18}}\right)^{\frac{3}{2}}+{\Xi_0}}\right)\nonumber\\
& & \times\left(\frac{1}{\sqrt{\tau_s+\mu_3{\cal V}^{\frac{1}{18}}}\left(\left(\tau_s+\mu_3{\cal V}^{\frac{1}{18}}\right)^{\frac{3}{2}}+{\Xi_0}\right)}+\frac{\tau_s+\mu_3{\cal V}^{\frac{1}{18}}}{\left(\left(\tau_s+\mu_3{\cal V}^{\frac{1}{18}}\right)^{\frac{3}{2}}+{\Xi_0}\right)^2}\right)\nonumber\\
& & -2n^s\sqrt{\tau_s+\mu_3{\cal V}^{\frac{1}{18}}}\left(\left(\tau_s+\mu_3{\cal V}^{\frac{1}{18}}\right)^{\frac{3}{2}}+{\Xi_0}\right)\left(-n^s+\frac{\sqrt{\tau_s+\mu_3{\cal V}^{\frac{1}{18}}}}{\left(\tau_s+\mu_3{\cal V}^{\frac{1}{18}}\right)^{\frac{3}{2}}+{\Xi_0}}\right)^2
\Biggr]
\end{eqnarray}
 Taking ${\cal V}\sim 10^6$, one can approximate ${\cal V}^{\pm\frac{1}{36}}\approx {\cal O}(1)$. Using
  (\ref{eq:extr axions}), one can approximate $\Xi_0\sim{\cal V}$. One can then verify from (\ref{eq:dV}) that $V^\prime(\tau_S)=0, V^{\prime\prime}(\tau_S)>0$ for
$\tau_S\sim{\cal V}^{\frac{1}{18}}\sim ln {\cal V}, n^s\sim{\cal O}(1)$ having already stabilized $\tau_B$ at ${\cal V}^{\frac{2}{3}}$. One can obtain the same result by a more careful calculation without assuming, to begin with, that the large divisor volume has been stabilized at ${\cal V}^{\frac{2}{3}}$.

\end{document}